 \documentclass{aa}  
\usepackage{graphicx}
\usepackage{txfonts}
\usepackage[utf8]{inputenc}
\usepackage{adjustbox}
\usepackage[usenames, dvipsnames]{color}
\usepackage{todonotes}
\usepackage{natbib}
\usepackage{paralist}
\usepackage{regexpatch}
\usepackage{tabularx}
\newcolumntype{L}[1]{>{\raggedright\arraybackslash}m{#1}}
\newcolumntype{C}[1]{>{\centering\arraybackslash}m{#1}}
\newcolumntype{R}[1]{>{\raggedleft\arraybackslash}m{#1}}
\usepackage{soul}

\usepackage{subfig}

\usepackage{lscape}
\usepackage{multicol}
\usepackage{graphicx}

\usepackage{tablefootnote}
\usepackage{threeparttable}

\usepackage{color}

\usepackage{url}

\def \teff {$T_{\mathrm{eff}}$}

\begin{document}

   \title{Catalog for the ESPRESSO blind radial velocity exoplanet survey \thanks{Based on observations collected at the La Silla Observatory, ESO(Chile), with the HARPS spectrograph at the 3.6-m telescope for program $97.C-0561(A)$, data obtained at the Paranal Observatory of the European Southern Observatory with UVES at VLT for program $097.C-0561(B)$ and data obtained at the TNG telescope for program $A33TAC\_7$, and also 95 programs. The large list is presented in Table~\ref{result} } } 

   \subtitle{}

   \author{S. Hojjatpanah \inst{\ref{IA}, \ref{U-Porto}}
   \and
    P. Figueira\inst{\ref{ESO}, \ref{IA}} \and
    N.C. Santos  \inst{\ref{IA}, \ref{U-Porto}} \and
    V. Adibekyan\inst{\ref{IA}} \and
    S. G. Sousa\inst{\ref{IA}} \and
    E. Delgado-Mena\inst{\ref{IA}} \and
    Y. Alibert \inst{\ref{bern}} \and
    S. Cristiani \inst{\ref{INAF-34143}, \ref{INAF-34127}} \and
    J.~I. Gonz\'alez Hern\'andez \inst{\ref{tenerife}, \ref{tenerife-uni}} \and         
    A. F. Lanza \inst{\ref{INAF-95123}} \and
    P. Di Marcantonio\inst{\ref{INAF-34143}} \and
    J. H. C. Martins \inst{\ref{IA}} \and
	G. Micela\inst{\ref{INAF-90134}} \and
	P. Molaro \inst{\ref{INAF-34143}} \and
	V. Neves\inst{\ref{brazil-1},\ref{brazil-2}} \and
    M. Oshagh \inst{\ref{gottingen},\ref{IA}} \and
    F. Pepe \inst{\ref{geneva}} \and
	E. Poretti\inst{\ref{INAF-23807},\ref{INAF-38712}} \and
	B. Rojas-Ayala\inst{\ref{u-chile}} \and
	R. Rebolo \inst{\ref{tenerife}, \ref{tenerife-uni}, \ref{CSIC}} \and
	A. Su\'{a}rez Mascare\~no\inst{\ref{geneva}} \and
	M. R. Zapatero Osorio\inst{\ref{madrid}}
          }
   \institute{Instituto de Astrof\'isica e Ci\^{e}ncias do Espa\c co, Universidade do Porto, CAUP, Rua das Estrelas, 4150-762 Porto, Portugal\label{IA}
	\email{Saeed.Hojjatpanah@astro.up.pt}     
\and
	Departamento de Fisica e Astronomia, Faculdade de Ci\^{e}ncias, Universidade do Porto, Rua Campo Alegre, 4169-007 Porto, Portugal \label{U-Porto}
\and
	European Southern Observatory, Alonso de Cordova 3107, Vitacura, Santiago, Chile \label{ESO}
\and
	Physikalisches Institut \& Center for Space and Habitability, Universit\"at Bern, Gesellschaftsstrasse 6, 3012 Bern, Switzerland  \label{bern}
\and
    INAF - Osservatorio Astronomico di Trieste, Via G. B. Tiepolo 11, Trieste, I–34143, Italy \label{INAF-34143}
\and
    INFN-National Institute for Nuclear Physics, via Valerio 2, I-34127 Trieste, Italy \label{INAF-34127}
 \and
 	Instituto de Astrof\'isica de Canarias, V\'ia L\'actea S/N, E-38205 La Laguna, Tenerife, Spain \label{tenerife}
\and
    Universidad de La Laguna (ULL), Departamento de Astrofísica, E-38206 La Laguna, Tenerife, Spain \label{tenerife-uni}
\and
    INAF-Osservatorio Astrofisico di Catania, Via S. Sofia, 78, 95123 Catania, Italy \label{INAF-95123}
\and
    INAF - Osservatorio Astronomico di Palermo, Piazza del Parlamento 1, 90134 Palermo, Italy \label{INAF-90134}
\and
	Instituto Federal do Paran\'a, 85860000, Campus Foz do Igua\c cu, Foz do Igua\c cu-PR, Brazil \label{brazil-1}
\and
	Casimiro Montenegro Filho Astronomy Center, Itaipu Technological Park, 85867-900, Foz do Igua\c cu-PR, Brazil \label{brazil-2}
\and
    Institut f\"ur Astrophysik, Georg-August-Universit\"at, Friedrich-Hund-Platz 1, 37077 G\"ottingen, Germany \label{gottingen}
\and
	Observatoire Astronomique de l'Universit\'{e} de Gen\`{e}ve, 1290 Versoix, Switzerland \label{geneva}
\and
	INAF--Osservatorio Astronomico di Brera, Via E.~Bianchi 46, 23807 Merate (LC), Italy \label{INAF-23807}
\and
	Fundaci\'on Galileo Galilei -- INAF, Rambla Jos\'e  Ana Fernandez P\'erez 7, 38712 -- Bre\~na Baja, Spain \label{INAF-38712}
\and 
	Departamento de Ciencias Fisicas, Universidad Andres Bello, Fernandez Concha 700, Las Condes, Santiago, Chile \label{u-chile}
\and
{Consejo Superior de Investigaciones Científicas (CSIC) \label{CSIC}}
\and
Centro de Astrobiolog\'ia (CSIC-INTA), Carretera Ajalvir km 4, E-28850 Torrejon de Ardoz, Madrid, Spain \label{madrid}
}

   \date{Received ----, ----; accepted --- , ----}

 
  \abstract
   {}
  {One of the main scientific drivers for ESPRESSO,\'Echelle SPectrograph, is the detection and characterization of Earth-class exoplanets. With this goal in mind, the ESPRESSO Guaranteed Time Observations (GTO) Catalog identifies the best target stars for a blind search for the radial velocity (RV) signals caused by Earth-class exoplanets.

 }
   {Using the most complete stellar catalogs available, we screened for the most suitable G, K, and M dwarf stars for the detection of Earth-class exoplanets with ESPRESSO. For most of the stars, we then gathered high-resolution spectra from new observations or from archival data. We used these spectra to spectroscopically investigate the existence of any stellar binaries, both bound or background stars. We derived the activity level using chromospheric activity indexes using $log(R' _{HK})$, as well as the projected rotational velocity $\textit{v sin i}$. For the cases where planet companions are already known, we also looked at the possibility that additional planets may exist in the host's habitable zone using dynamical arguments.}
   {We estimated the spectroscopic contamination level, $\textit{v sin i}$, activity, stellar parameters and chemical abundances for 249 of the most promising targets. Using these data, we selected 45 stars that match our criteria for detectability of a planet like Earth. The stars presented and discussed in this paper constitute the ESPRESSO GTO catalog for the RV blind search for Earth-class planets. They can also be used for any other work requiring a detailed spectroscopic characterization of stars in the solar neighborhood.
   }
  {}

   \keywords{
                Planetary systems, Planets and satellites: composition, Techniques: radial velocities, spectroscopy, Stars: abundances, activity
               }

   \maketitle
%

\section{Introduction}

  Since the discovery of 51 Peg b \citep{mayor1995jupiter}, the radial velocity (RV) method has been extensively used to detect, confirm, and estimate the planetary mass of exoplanets.
  Over the past years, hundreds of exoplanets have been discovered with the RV technique using diverse spectrographs with different levels of precision in the measurement of RVs. For instance, HARPS  (High Accuracy Radial velocity Planet Searcher), on the European Southern Observatory 3.6m telescope at La Silla, is one of the most successful planet-hunters. It has achieved an unprecedented precision of about $\rm 80 \;cm \;s^{-1}$ over long observing periods \citep{pepe2014instrumentation}.
After 15 years of intensive observations and many groundbreaking results \citep[e.g.,][]{santerne2018earth,lovis2011harps7planet,anglada2016terrestrial,diaz2016harps} HARPS has arguably reached its RV precision limit, and a new spectrograph was designed to be its successor: ESPRESSO (\'Echelle SPectrograph for Rocky Exoplanets and Stable Spectroscopic Observations) the new ultra-stable, fibre-fed, cross-dispersed, high-resolution \'echelle spectrograph operates at the combined coud\'e focus of the European Southern Observatory's Very Large Telescope (VLT) \citep{pepe2014espresso, 2017arXiv171105250G}.\\

It is important to note that ESPRESSO was designed to target several science drivers, including exoplanets, variations of fundamental constants, and spectroscopy of extra-galactic stars \citep{pepe2010espresso}. Among others, ESPRESSO will also allow many other problems of modern astrophysics to be tackled, including the measurement of stellar oscillations, the derivation of detailed chemical abundances in stars, as well as the study of interstellar gas.\\
  
ESPRESSO is able to use the superior light collection capabilities of ESO's VLTs and reach a RV precision of $\rm 10 \;cm \;s^{-1}$ on bright nearby stars in a matter of minutes. Such a value is crucial if we want to detect the signature of an Earth-class planet orbiting inside the habitable zone of a main-sequence, solar-type star. The expected RV modulation amplitude for the Earth around the Sun is at most about 9 $\;cm \;s^{-1}$.\\

The main scientific driver for ESPRESSO is the measurement of high-precision RVs of G, K, and M dwarfs to search for rocky planets inside the habitable zone (HZ). Eighty percent of the observing nights in the Guaranteed Time Observations (GTO) program will be dedicated to exoplanet science, having in mind the detection of rocky planets inside the HZ, the characterization of low mass planets discovered using high-precision photometric transits, and the characterization of exoplanet atmospheres. The ESPRESSO GTO catalog presented in this paper is meant to tackle the first of these cases by selecting the best candidates for detecting RV signals of low mass planets.\\
To obtain the most precise RV measurements, several criteria should be considered to select the best targets. We defined a photon noise criterion in order to select the brightest stars. This criterion is a conservative approximation of the achievable photon noise by ESPRESSO. We also determined the activity indexes $log(R' _{HK})$ and $H_{\alpha}$, the projected rotational velocity ($\textit{v sin i}$), and the signature of the spectroscopic contamination for 249 bright and nearby stars. This is a new, unique data-set that is of natural interest to a wide range of spectroscopic studies of G, K, and M type solar neighbors. By using focused criteria, we selected the most visible and bright stars for ESPRESSO. We also derive precise stellar parameters and chemical abundances for most of our targets.\\

The outline of this paper is as follows. We first describe our data and new observations in Sect.~\ref{Observations and Data}. We investigate the companion signature in Sect.~\ref{binary}. The procedure to determine activity indicator and $\textit{v sin i}$ is explained in Sects.~\ref{activity} and \ref{vsini}, respectively. Targets with already known planets are considered in Sect.~\ref{known planet}. We explain stellar characterization and chemical abundance in Sect.~\ref{stellarparam}. We investigate the rotational period of M dwarfs in Sect.~\ref{period}. Our summary and conclusion are presented in Sect.~\ref{sum}.

\section{Observations and data \label{Observations and Data}} 

\subsection{Target selection \label{Catalog and target selection}}

The first step to define the sample of stars was to identify the brightest G, K, and M stars in the sky. Since there is no comprehensive catalog for solar neighborhood G, K, and M dwarfs, we decided to use two different catalogs to determine initial samples. The two catalogs used were \textit{Hipparcos} for G and K stars by \citet{perryman1997hipparcos}, and \citet{2011AJ....142..138L}  for M stars. The \textit{Hipparcos} catalog is very probably complete down to the G and K faintest stars of interest; and \citet{2011AJ....142..138L} was selected as the most complete catalog for M dwarfs. This is an all-sky catalog of M-dwarfs with apparent infrared magnitude J < 10 that provides parallax measurements and photometric distance. \citet{2011AJ....142..138L} notes that the catalog is not absolutely complete, reporting completeness levels of $\approx$ 75$\%$ bright M dwarfs (J < 10), which cover $\approx$ 60$\%$ of the southern sky. However, it is the most comprehensive and suited for our goal.

At the outset, we selected all stars from the \textit{Hipparcos} catalog with declination < +30 deg (i.e., well visible from Paranal). We recovered their V and K apparent magnitudes by matching \textit{Hipparcos} to the 2MASS catalog \citep{cutri2003vizier}. We used Eq. 6 in \citet{casagrande2008m} and Eq. 3 in \citet{casagrande2010absolutely} for the color calibrations and estimated the stellar effective temperatures (\teff). We determined an approximate spectral type from \teff\ using \citet{carroll2017introduction}. The goal was to classify the spectral type and select G, K, and M stars for the primary sample. We attributed spectral types based on the spectroscopic after the stellar characterizations in Sect.\ref{stellarparam} (as seen in the Table~\ref{result}).\\
 
 To the stars mentioned above we added stars with confirmed exoplanet(s) using all methods as discussed in Sect.~\ref{known planet}. Around 22 \% of the known exoplanet systems have more than one planet detected corresponding to 634 multi-planet confirmed systems by November 2018\footnote{\texttt{https://exoplanetarchive.ipac.caltech.edu}}. There are also 817 multiplanet system candidates detected by NASA's Kepler spacecraft and certainly more will be identified by TESS, Transiting Exoplanet Survey Satellite, soon \citep{ricker2014transiting}. The high occurrence rate shows these systems are common. As an additional factor of interest for a given host, we considered the possibility of the presence of an Earth-mass planet inside the HZ of a known (i.e., confirmed) planet system (see Sect. \ref{known planet}).

\subsection{Cleaning by applying a photon noise criterion}

For an Earth-mass exoplanet inside the HZ to be detectable by ESPRESSO, the RV signal created by the planet  should be larger than the photon noise of our measurements. To derive the location of the HZ around each star of the sample, we used the definition of \citet{selsis2007habitable}, which satisfies two general conditions for HZ: 1) there should be the capability of hosting liquid water on the surface at any temperature between 273 K (triple point of water) and 647 K(critical temperature of water); and 2) the amount of $CO_{2}$ in the planet's atmosphere should be abundant enough when the mean surface temperature decreases to less than 273 K (for more information see \citet{selsis2007habitable}).
We determined the position of the continuous HZ using the definition from \citet{selsis2007habitable} for the "extreme theoretical limits of the continuous habitable zone" and of an RV semi-amplitude signal ($K_{max} = \rm 10 \;cm \;s^{-1}$) for $1_{\sigma}$. We then computed $K_{max}$ for an Earth-mass planet at a semi-major axis corresponding to the inner boundary of the HZ using Eq. 2.28 in \citet{perryman2011exoplanet}:
  
  \begin{equation}
  \label{k-formul}
  K = 28.4 ms^{-1} \left(\frac{P_{p}}{1 yr} \right)^{-1/3}  \left(\frac{M_{p} \sin i}{M_{J}}\right) \left(\frac{M_{\star}}{ M_{\sun}}\right)^{-2/3},
  \end{equation}
  where $P_{p}$, $M_p \sin i $, are the period and minimum mass of the planet, and $M_{\star}$ is the mass of the host star.
  We consider the inner boundary in order to provide the optimistic case with the maximum $K$ value.
  
  \begin{table*}
\tiny
\caption{RV signal semi-amplitude of one Earth-mass planet for HZ limits}             
\label{table:rv-signal}      
\centering                          
\begin{tabular}{c c c c c c}        
\hline\hline                 
Spectral Type  & Mass[$M_{\sun}$] & $a_{min}[AU]$ & $a_{max}[AU]$ & $K_{min}[cm~s^{-1}]$ & $K_{max}[cm~s^{-1}]$  \\    
\hline                        
  G0 & 1.0 & 0.5 & 3.0 & 5.2 & 13 \\      
  G5 & 0.9 & 0.4 & 2.0 & 6.7 & 15   \\
   K0 & 0.8 & 0.3 & 1.5 & 8.2 & 18 \\
   K5 & 0.7 & 0.2 & 1.0 & 11 & 24 \\
   M0 & 0.5 & 0.1 & 0.5 & 18 & 40 \\
   M5 & 0.2 & 0.04 & 0.2 & 45 & 100 \\
   M8 & 0.1 & 0.01 & 0.08 & 100 & 283 \\
   
\hline      
\end{tabular}
\end{table*}
  
  HARPS is able to reach a precision of about $\rm 1 \;m \;s^{-1}$ in one minute on a late G dwarf of magnitude $V$=7.5. This project was started several years ago, much before the first ESPRESSO data were obtained. We used HARPS at the 3.6m as a starting point for the photon noise calculations of ESPRESSO at the VLT. Due to the larger collecting area (i.e., mirror sized of 8.2-m vs. 3.6-m), we assumed a gain of 1.75 magnitude for ESPRESSO relative to HARPS. Therefore we considered that ESPRESSO will have a gain of a factor of five in flux, and that the measurements would be done with a minimum integration time of 15 minutes \citep[to average out oscillation modes][]{dumusque2011planetaryb}. We calculated the photon noise achievable by ESPRESSO by scaling it from experience using HARPS. As a consequence, the photon noise limits for $V$ can be defined by,
 \begin{equation}
  V = 5\log(K_{max} \;cm \;s^{-1}) + 2.1.,
\end{equation}
where $V$ is an estimation for the visual magnitude limit. By application of this formula to the $K_{max}$ in Table~\ref{table:rv-signal}, we estimated the following cutoff as the photon noise criteria: G0: $V$ < 7.7, G5: $V$ < 8.0, K0:$V$ < 8.4, K5: $V$ < 9.0, M0: $V$ < 10.1, M5: $V$ < 12.1, and M8: $V$ < 14.4.\\
These values are compatible with the RV precision, calculated using the ESPRESSO Exposure Time Calculators, ETC\footnote{https://www.eso.org/observing/etc/}. For example, using a K2V template for a star in magnitude 8.4 and exposure time of 15 minutes, we have $\rm 19 \;cm \;s^{-1}$ for the RV photon noise precision. Using the M2V template available in ETC, for a star in magnitude 12.1 and exposure time of 15 minute, we have $\rm 110 \;cm \;s^{-1}$ for the RV photon noise precision. We should point out that such an estimation for the photon noise is heavily dependent on spectral type. Most of the lines that have been used to measure RV are in the blue part of the spectrum for G spectral type and in the red part for the M spectral types. As such the precision is not only dependent on the visual magnitude, and here we only considered photon noise limitation \citep{nealeniric}.

\subsection{Photometric variability in Hipparcos catalog}

In the \textit{Hipparcos} catalog, we considered large photometric variability as an exclusion criterion and discarded stars with a variability flag larger than 0.06 mag. We considered stars with \textit{HvarType} flags C, M, R, U \footnote{In Hipparcos: Hipparcos-defined type of variability flag that were C, M, R, and U (constant, micro-variable, revised color and unknown)} and also stars without any \textit{HvarType} flags in the catalog. We identified a target as a single component and isolated star by using the \textit{Ncome} \footnote{Hipparcos: Number of components in this entry.} flag and by matching \textit{Hipparcos} to the 2MASS catalog for the infrared counterpart of these stars as the closest source within 5" of the optical coordinates.
 
 \subsection{Surface gravity}

In the next step we estimated the surface gravity of each star, with the goal of excluding evolved stars. Several studies have shown that for evolved stars the RV's jitter is much larger than for main sequence stars \citep[e.g][]{setiawan2004precise,hatzes2005giant,hekker2008precise,2007A&A...472..657L}.
From Eq. 1 of \citet{santos2004spectroscopic}, using the mentioned quantities and the \textit{Hipparcos} parallax, we calculated the $\log g$ of the stars from 
 
  \begin{equation}
  \label{logg_hip}
  \begin{array}{rl}
  
  \log (\frac{g}{g_{\sun}}) = &\log(\frac{M}{M_{\sun}})  + 4 \log(\frac{T_{eff}}{T_{eff\sun}}) + \\[1em]
  & 2 \log \pi + 0.4 (V + BC) + 0.11 
  \end{array}
\end{equation}

Where \textit{$V$} is the visual magnitude, \textit{$\pi$} is the parallax in arcseconds, \textit{BC} is the bolometric correction, and \textit{M} is the stellar mass. Using each \teff\ we attributed an approximated mass at main-sequence, and bolometric correction to each star \citep{carroll2017introduction}.
Due to a miscalculation in $\log g$ used in the selection process, we selected a large number of stars with low $\log g $ for further analysis. These were discarded in a later stage when a precise $\log g$ was derived from the spectra, having no impact on the final results. We applied a cutoff on $\log g$ > 4.1 $cm~s^{-2}$ to exclude evolved stars from the final sample.
 The list of stars fulfilling all our criteria in \textit{Hipparcos} has 187 entries, from which 15 were identified as spectroscopic binaries by \citet{pourbaix2004s}, and were discarded (see Table~\ref{sp-bin}). Of the remaining 172 stars, we identified 70 stars that had been observed with HARPS or UVES at least two times with a signal-to-noise ratio ($S/N$) of at least 50 for G and K, and 10 for M spectral type. The acquisition of new observations is discussed in Sect.~\ref{new-observation} (see Table~\ref{table:data}). \\

  \subsection{Sampled M-dwarfs}

 We repeated the procedure described in Sect. \ref{Catalog and target selection} for the M-dwarf catalog prepared by \citet{2011AJ....142..138L}. Most of the stars in this catalog are closer than 60 parsec and are of K7 to M4 spectral type, except for a few late M-dwarfs. We searched for stars respecting the $V$ < 14.5 and declination < +30 deg criteria. We followed the same procedure while ignoring the $\log g$ restrictions, as in this dataset giants are supposed to have already been excluded (but see Sect. \ref{screening-logg}). By applying these criteria, we selected 114 M stars from \citet{2011AJ....142..138L}.

 Fifty-two of the stars selected above have high-quality data in the ESO archive, six of the stars are listed as spectroscopic binaries by \citet{pourbaix2004s}\ (see Table~\ref{sp-bin}) and five others were identified as binaries or as having close companions using Simbad database \footnote{\texttt{SIMBAD Astronomical Database: http://simbad.u- strasbg.fr/Simbad/}: PMJ13550-2905, PMJ16554-0819, PMJ16554-0820, PMJ16589-3933N, PMJ17464+2743W} records. The remaining forty-one stars were observed (see Sect.~\ref{new-observation} and Table~\ref{table:data}).\\

 \subsection{Screening the M-dwarfs in the sample by $\log g$ \label{screening-logg}}
 
\begin{figure}
   \centering
   \includegraphics[width=9.3cm]{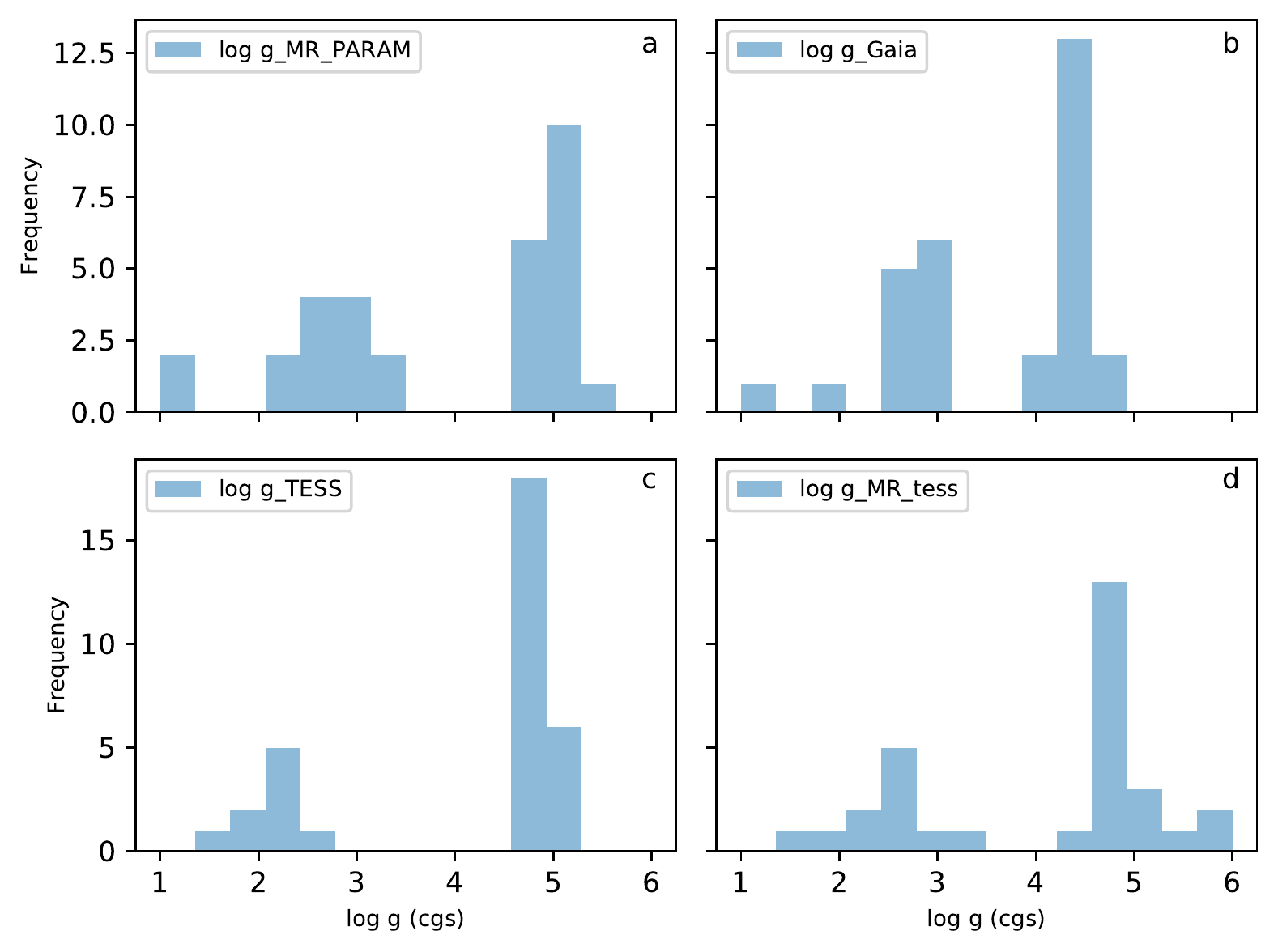}

      \caption{Distribution of  $\log g $ for cooler stars in final sample. This was completed using: a) mass and radius derived by theoretical isochrones; b) Gaia parallaxes; c) directly from TIC; and d) mass and radius reports in TIC.}
         \label{logg-compar}
   \end{figure}

 As mentioned above, the M-stars in our sample were supposed to be all dwarfs. However, to check for this assumption, we derived $\log g$ values for all the stars using the following approaches:
 
 \begin{itemize}
  \item $\log g$ using theoretical isochrones:
  
We used the theoretical isochrones described in \citet{da2006basic} and the Padova web interface\footnote{\texttt{http://stev.oapd.inaf.it/cgi-bin/param$\_$1.3}}. We then used the mass and radius derived with this method in order to determine the $\log g$ (Fig.~\ref{logg-compar}-a), using the formula directly:

\begin{equation}
\label{mrg}
\log g =  \log_{10}\left(\frac{GM}{R^{2}}\right).
  \end{equation}
  
  \item $\log g$ using Gaia parallaxes (trigonometric gravities):

  After the Gaia Data Release 2 \citep{brown2018gaia}, we derived $\log g$ by using the Eq. \ref{logg_hip} with Gaia parallax values (Fig.~\ref{logg-compar}-b). We used the mass values in TIC\footnote{Tess Input Catalog} and the BC extracted using Table 1 in \citet{torres2010use} as a function of temperature.
  
  \item $\log g$ in TESS Catalog:

After the TESS input catalog release, we identified the targets in the TIC and used the $\log g$ in the catalog \citep{stassun2018tess} (Fig.~\ref{logg-compar}-c). We also calculated $\log g$ by mass and radius reports in TIC and Eq. \ref{mrg} (Fig.~\ref{logg-compar}-d).
\end{itemize}

In the Fig.~\ref{logg-compar}, we present the distribution of the $\log g$ for cool stars (\teff\ < 3950) as derived by each method. There are ten stars\footnote{PMJ03219+0940, PMJ03340+1123, PMJ03377+1042, PMJ16015-4710, PMJ16051-0508, PMJ17173+0003, PMJ18400-1644, HIP78508, HIP79148, HIP81705} in the \citet{2011AJ....142..138L} that are characterized as having $\log g$ < 3.5 by all the above methods. As such, these are confidently classified as giant stars; we excluded these stars from the final list. There are also five stars\footnote{PMJ15357-2812, PMJ15543-3536, PMJ16046-5521, PMJ16118-3331, PMJ19473-2424} for which the $\log g $ have different values using the Gaia parallax and TIC if available. Therefore, these targets were flagged as potential giants and will be followed with high caution when performing RV measurements.\\
 
 \subsection{Acquisition of new high-resolution observations} \label{new-observation}

High-resolution spectra for the previously described list of 122 stars for which no high-resolution high-$S/N$ data were available, were requested in open time calls of an ESO proposal (program ID: 097.C-0561(A-B)\footnote{PI: P. Figueira}), and then through a HARPS-N complementary proposal (program ID: A33TAC\_7\footnote{PI: P. Molaro}). The instruments used were HARPS, HARPS-N \citep{2012SPIE.8446E..1VC} mounted on the 3.58m Telescopio Nazionale Galileo at Roque de los Muchachos Observatory in La Palma (Canary Islands, Spain), and UVES \citep{2000SPIE.4008..534D} spectrographs at Unit Telescope 2 of the VLT array. \\

Since M stars are faint in the optical, most of the M dwarfs identified as potential candidates were observed with UVES, mounted on the VLT. The choice of UVES as an instrument provides, just like HARPS, full coverage of the spectral optical domain, allowing the activity indicators to be covered. As an example, for a $V$=12.0, M2 star we obtained a $S/N$ of $\sim$ 15 at the Ca II H $\&$ K regions, and $S/N$ $\sim$100 at 600nm central wavelength.

\begin{table}

\caption{Summary of Data sources and number of spectra}             
\label{table:data}      
\centering 
\small
\tiny
\begin{tabular}{c c c c c c}        
\hline\hline                 
 & G and K type & M type & HARPS & HARPS-N & UVES  \\    
\hline                        
  Archive & 28 & 95 & 100 & 0 & 22 \\      
   New Observation & 78 & 44 & 78 & 44 & 42   \\
   Total & 106 & 139 & 178 & 44 & 64 \\
   
\hline                                   
\end{tabular}
\end{table}

 The final sample consists of stars observed from 3 to 8 May 2016 using HARPS, UVES UT2-Kueyen between April-July 2016, and using HARPS-N, between July-September 2016. For each target, at least two spectra were collected on different nights. 
 In Table~\ref{table:data}, a brief overview of the observations is presented. The average $S/N$ for spectra obtained by HARPS and HARPS-N is 210 and by UVES is 51 at mid-wavelength (around 550 nm).\\

\subsection{Data Reduction and Cross-Correlation Function derivation \label{data reduction and ccf}}

For HARPS and UVES, we collected all reduced, raw, and associated calibration-reduction spectra taken each night using the ESO archives\footnote{\texttt{$https://www.eso.org/sci/observing/phase3/data$\_$streams.html$}}. We performed the same data collection procedure for the 122 targets that were already observed at least twice using HARPS and UVES spectrographs, above the threshold of 210 for HARPS and HARPS-N and 51 for UVES. For HARPS-N and HARPS spectra data reduction, we used the latest version of the HARPS-N pipeline (DRS V3.8) using the Cross-Correlation Function, CCF technique with a template of G2, K5, and M2 masks \citep{pepe2002coralie,baranne1996elodie}. This pipeline provides the CCF's full width at half maximum (FWHM), CCF's bisector span inverse slope, and the Ca II activity index S and $Log(R' _{HK})$ for F, G, and K stars.

In order to evaluate the presence of a second set of spectroscopic lines, we need to calculate the CCF in a wide window (see Sect.~\ref{binary}). We used the following procedure to calculate the CCFs in a range of $\rm 200 \;km \;s^{-1}$ to ensure a contaminant CCF peak would not remain unnoticed. \\

The UVES data were reduced by a custom-made Python script that interfaces directly with ESO's CPL (Common Pipeline Library) data reduction recipes for the UVES spectrograph \citep{Martins2017}. This method permits two different approaches: i) run the \textit{red chain} recipe
that performs the whole reduction; and ii) perform a step-by-step data reduction using the individual data reduction recipes. We selected the latter approach as it allows finer control over the data reduction process.

It should be noted that when performing the flat-fielding of the science raw spectra, the UVES reduction recipes will remove the instrumental profile (i.e., \textit{blaze}) information from the spectra. This means that the reduced spectra loses the information about the real flux, not allowing us to estimate the noise level at each wavelength. Due to this effect, the correction of the blaze function will give incorrect weights to the spectral lines closer to the edges when computing the CCF. To tackle this issue, we used the extracted, reduced, master-flat spectra for each order to reconstruct the \textit{blaze} function, meaning the instrumental profile of the instrument on each order. These were constructed by performing a moving average over the extracted flat field for each order with a window of 100 pixels in the spectral space. This allows us to recover the instrumental profile, by smoothing the flat function over a 100-pixel region in the spectrum space to discard local defects (e.g., bad pixels). Each two-dimensional spectrum was then multiplied by the reconstructed \textit{blaze} function to recalculate the spectra with the real flux.
		
To compute the CCF of each individual spectrum, we used the spectral masks from the HARPS DRS \citep{2003Msngr.114...20M}. It should be noted that although they were originally built to be used with HARPS data, the wavelength coverage of both instruments is similar enough ([450-750]nm for UVES, against [378-691]nm for HARPS) that the same masks can be used. For each individual spectrum order and a given radial velocity $RV_i$, our implementation of the CCF is mathematically defined by 

\begin{equation}
  CCF_{order}(RV_i) = \sum_{lines,order} \left( d_{line} \times \sum_{pixels} \left(f_{line,pix} \; F_{line,pix} \right)    \right),
\end{equation}
                
where $d_{line}$ is the depth of the selected line (from the binary mask); $pixels$ represents the list of spectral pixels covered by the selected line; $f_{line,pix}$ the fraction of the pixel that falls on the line; and $F_{line,pix}$ the flux of the line for each pixel (from the spectrum). Note that the $line$ parameter is dependent on $RV_i$, as it refers to the spectral line wavelength position for the given $RV_i$. The CCF value is then computed over a RV range centered on the expected RV of the system, yielding a plot such as the one presented in Figure \ref{ccf-two-peak}.

\section{Signature of stellar companions in the CCF \label{binary} }

One of the goals of this study is to identify stars that are binaries or affected by the presence of a background star inside the fibre. For G and K type stars the presence of a contaminator can create a systematic RV signal of $\rm 10 \;cm \;s^{-1}$ even in the extreme case when the difference in magnitude is of ten; this parasite signal can be larger than $\rm 1 \;m \;s^{-1}$ if the difference in magnitude is less than eight \citep{cunha2013impact}.   
For M type stars, a difference of magnitude between target and contaminant of eight, creates an effect of $\rm 10 \;cm \;s^{-1}$ and a difference of magnitude between target and contaminant of six leads to an effect of $\rm 1 \;m \;s^{-1}$ \citep{cunha2013impact}. As such, a careful examination should be made to search for hints of companions. \\

\begin{figure}
  \centering
     \includegraphics[width=9.4cm]{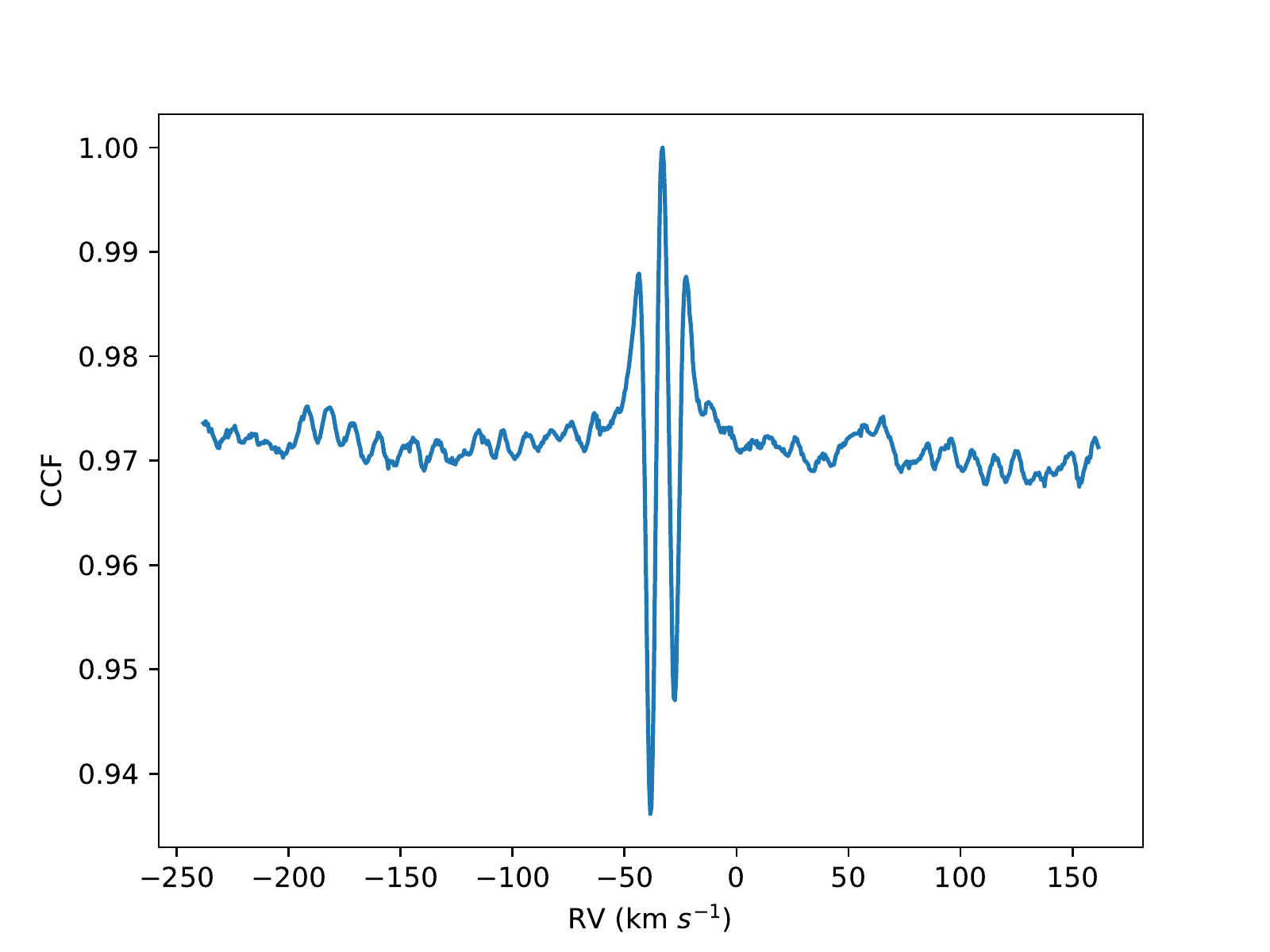}
      \caption{Example CCF of star with two significant peaks.}
         \label{ccf-two-peak}
  \end{figure}

There are several methods that can be used to check if the star has any detectable contaminations. Here we investigate using the CCF. The goal is to determine CCF's pollution induced by a second star in the sky, close enough to contribute with light to the total collected in the fiber. This companion can be gravitationally bound or fortuitously aligned with the star. If the companion or companions exist and are bright enough, the presence of a second set of lines can create either a secondary correlation peak, if well-separated in RV from the target, or a change in the CCF line profile, if blended with the target. 
After calculating the CCFs in a wide window ($\rm 200 \;km \;s^{-1}$), we compared the different CCFs obtained on each star for any anomaly in its usually "Gaussian shape" as well as of variability in its profile. In the first approach, we searched for the characteristic double peak signature, and identified six targets, five of which are known binaries with significant CCF contamination. We considered a Gaussian-like shape with a CCF depth of at least 10\% (an admittedly arbitrary value) to be a secondary peak. In Fig.~\ref{ccf-two-peak}, it is possible to see an example of double-line CCFs due to spectroscopic contamination. We include these targets with a signature of binary nature with "Y+" flag in Table~\ref{result}.

\begin{figure}
   \centering
  \includegraphics[width=9cm]{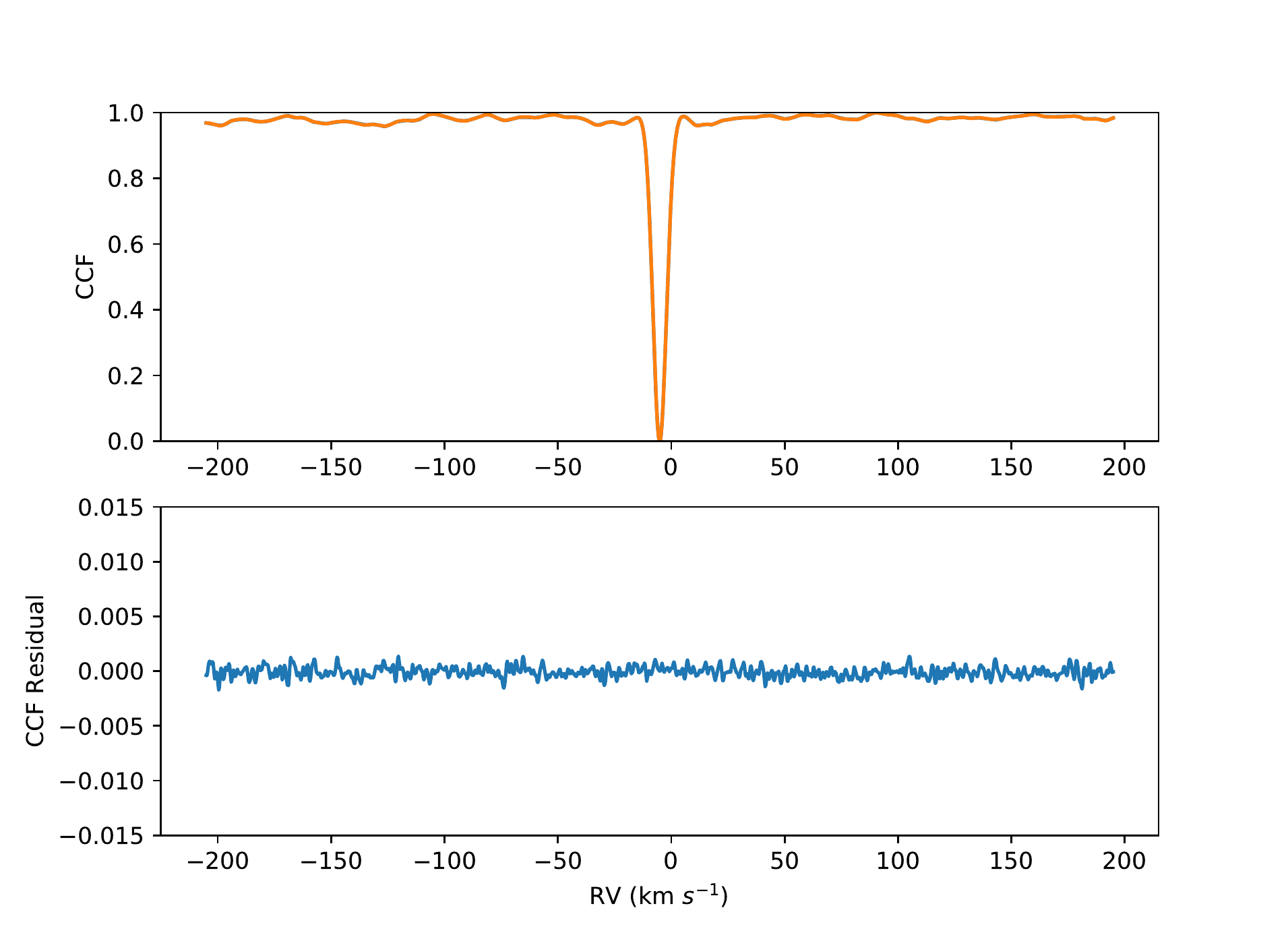}
     \includegraphics[width=9cm]{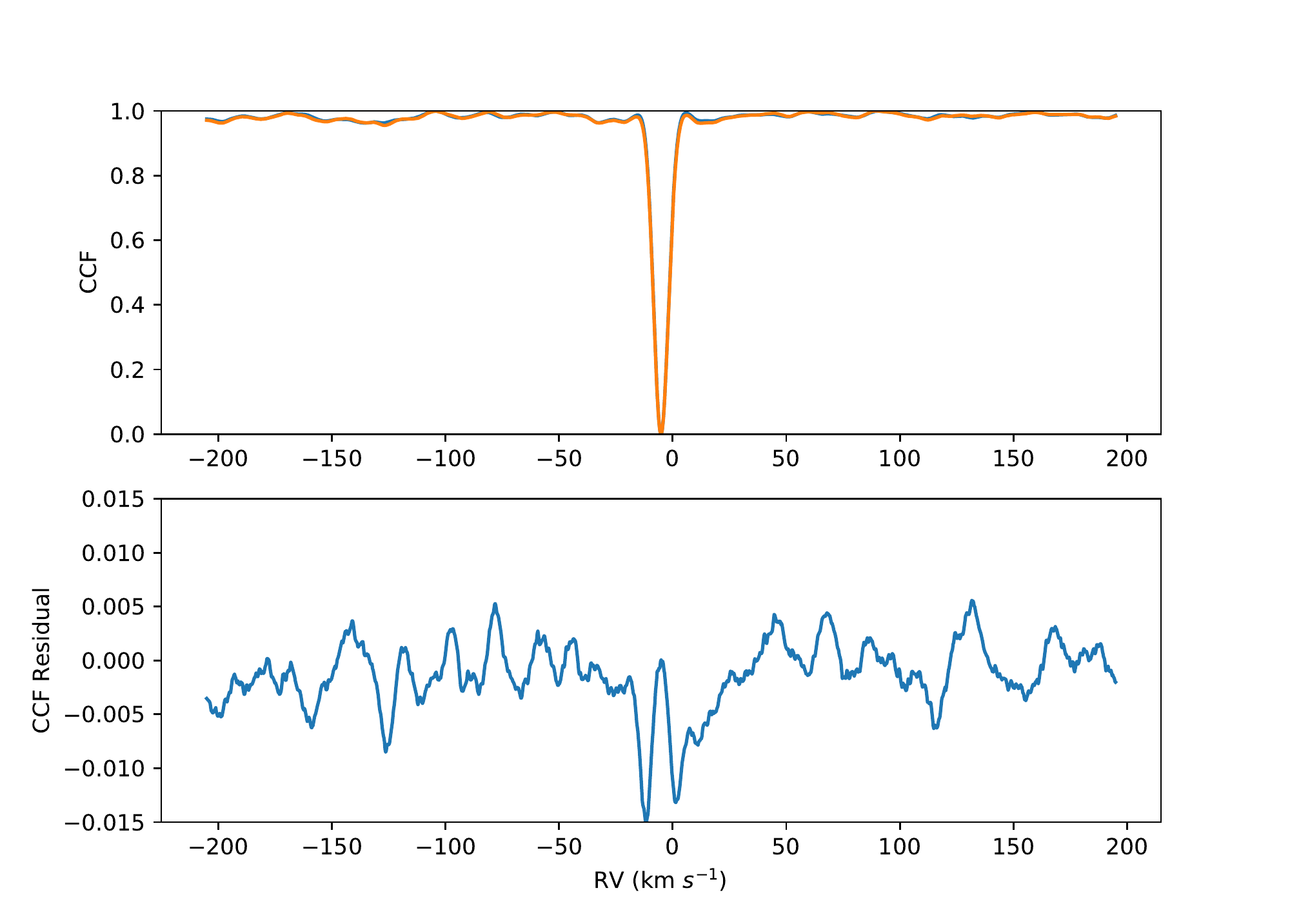}
      \caption{ Two examples of normalized CCFs and residuals of stars. Upper panels: without signs of contamination; bottom panels: with hints of contamination.}
         \label{resi-harps}
   \end{figure}

After this visual inspection procedure, we applied a more homogeneous criterion to identify stars on which a second component was present. We followed the method of \citet{2016A&A...585A..46B}: after the derivation of CCFs, we shifted the reduced CCFs to a common RV value and mutually subtracted them. We then searched if there was a significant peak in the CCF residuals (see Fig.~\ref{resi-harps}). The criteria for the absence of an additional peak is defined by

 \begin{equation}
 \label{snr_b}
   \sqrt{(S/N_{1})^{2}+(S/N_{2})^{2}} < \frac{1}{\sigma_{RMS}} .
   \end{equation}

Here the $S/N_{1}$ and $S/N_{2}$ are the median order signal-to-noise ratios of the CCFs, which are available in the reduced FITS headers; and $\sigma_{RMS}$ is the root mean square, RMS of the residual of the CCF difference. If the above condition is not met, then this means there is an additional source of scatter in the data. Equation~\ref{snr_b} provides a systematic way of searching for the secondary peak on CCFs affected by contamination, provided we have at least two spectra. Stars for which our analysis satisfies the above condition are flagged with "N", otherwise the target contamination flag is set as "Y". The results are presented in Table~\ref{result}.

In most cases, we did not find any significant hint of companion. Twenty-one stars have a companion and were discarded according to this criteria.

\section{Chromospheric activity indicator $log(R' _{HK})$  \label{activity} }

Stellar activity related phenomena are known to produce significant RV signals. These can both induce noise in our data or produce coherent signals that can mimic the signature of an orbiting planet. Stellar active regions (e.g., dark spots and bright faculae) produce RV signals at the level of exoplanet orbital signals when they combine with stellar rotation. Stellar active regions also deform the CCF profile. As a consequence, they generate parasite RV signals \citep{meunier2010using,dumusque2014soap,oshagh2018noise}. As such, it is mandatory to track and characterize activity signals in order to disentangle the real exoplanet RV signal \citep[e.g.,][]{hatzes2013investigation,queloz2001no,hatzes2013radial,santos2014harps,robertson2014stellar,suarez2017characterization,mascareno2018ropes}.

To clear our sample from active stars, we used an activity indicator, $Log(R' _{HK})$. We started by determining the S-index. This is a well-known quantity used for the first time in the long-term observations in the Mount Wilson stellar activity program \citep{1978PASP...90..267V,duncan1991ii}.

We used Ca II H $\&$ K lines and $R\&V$ continuum band-passes for S-index determination, 

 \begin{equation}
    S = \alpha \frac{N_{H }+ N_{K}}{N_{R} + N_{V}} \,.
   \end{equation}

The S-index measures the flux in the core of the line, normalized to the continuum band-passes. In this equation, $N_{H}$, $N_{K}$, $N_{R}$ and $N_{V}$ are the total flux in each band-pass and $ \alpha$ is a calibration constant usually adjusted at 2.30 \citep{duncan1991ii}. The 1.09 $\AA$ wide bands centred at 3933.664 $\AA$ and 3968.470 $\AA$ were used for K\&H respectively. For V\&R we used the 20 $\AA$ wide bands centred at 3901.070 $\AA$ and 4001.070 $\AA$, respectively.

To compare different stars' activity, we need to remove the effect of the photospheric contribution and perform a normalization depending on the color index ($B-V$) that was introduced. The $log(R' _{HK})$ has been defined by \citet{1984ApJ...279..763N}

 \begin{equation}
      R' _{HK}  = 1.34 \times 10^{-4} \cdot C_{cf}(B-V) \cdot S-R_{phot}(B-V) \,,
   \end{equation}where $C_{cf}(B-V)$ is the conversion factor \citep{1984ApJ...279..763N,middelkoop1982magnetic}. To derive $C_{cf}(B-V)$ we used the recent work of \citet{suarez2015rotation}, who calibrated these values in the range of B-V between 0.4 and 1.9:

 \begin{equation}
log_{10}C_{cf} = 0.668 - 1.270 \cdot B-V + 0.645 \cdot B-V^{2} - 0.443 \cdot (B-V)^{3}.
 \end{equation}\citet{suarez2015rotation} call photospheric contribution to the H  and K bandpasses $R_{phot}$ and we used the calibration defined as
 
 \begin{equation}
      R_{phot}   = 1.48 \cdot 10^{-4} \cdot exp[-4.3658 \cdot (B-V)].
  \end{equation}

\begin{figure}
   \centering
   \includegraphics[width=9cm]{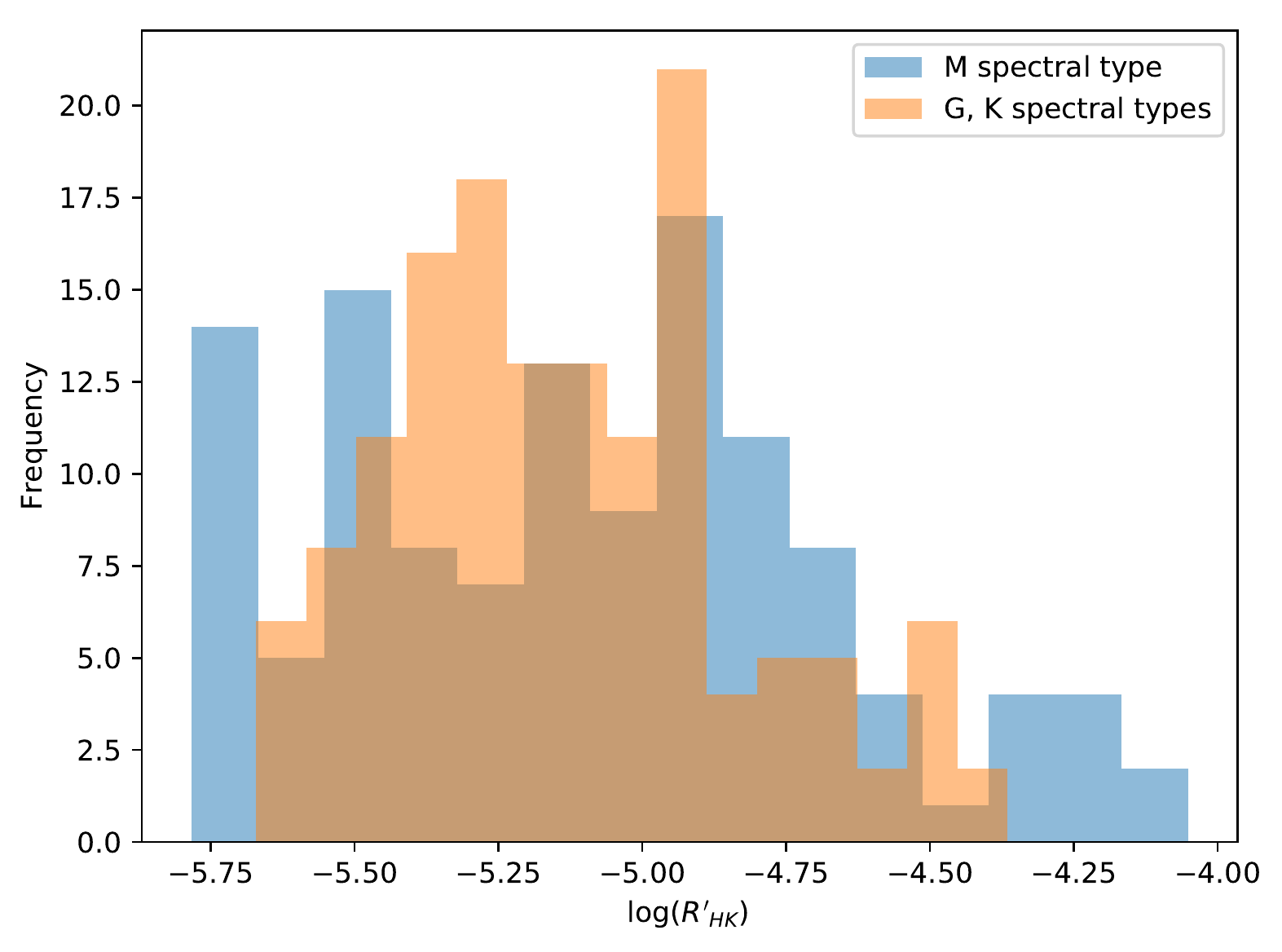}
    
      \caption{Derived $log(R' _{HK})$ distribution for star sample. }
         \label{logrhk}
   \end{figure}

Using these equations and measuring the fluxes in our spectra, we determined the S-index and $log(R' _{HK})$. The HARPS and HARPS-N DRS measure the S-index and convert this into the chromospheric activity indicator $log(R' _{HK})$. Both pipelines need $B-V$ and guess RV as input parameters. We used the $B-V$ values from the catalog data (e.g., Hipparcos) and the extracted RV from the CCFs.
For HARPS-N, we used YAbI\footnote{http://ia2.oats.inaf.it} a web based access to the pipeline of HARPS-N. The pipeline provided the calibrated S-index and $log(R' _{HK})$ for each spectrum in a similar way to HARPS.\\
For the stars observed with the UVES spectrograph, we corrected the RV shift using the RV's determined by the CCFs and then calculated the $log(R' _{HK})$. \\
We identified 43 stars with chromospheric activity higher than $-4.80$ dex, which usually is considered as the threshold for active stars \citep[e.g.,][]{vaughan1980survey}. On the other hand, stars less active than this value are usually seen as stars with solar-like activity. We removed those active stars from our final sample. The derived activity values for all the stars are presented in Table~\ref{result} and illustrated in Fig.~\ref{logrhk}.
 
\section{Projected rotational velocity ($\textit{v sin i}$)}\label{vsini} 
\label{vsini-section}

Another important parameter affecting RV precision is the projected rotational velocity of the host star ($\textit{v sin i}$). Assuming, for simplicity, that the spectral lines are resolved and have a Gaussian shape, RV precision is inversely proportional to the square root of the FWHM of the lines \citep[e.g.,][]{bouchy2001fundamental, figueira2018deriving}. For a non-resolved spectral line, $\sigma_{\rm RV} \propto (v \sin i)^{1.5}$, if the other broadening mechanisms can be neglected in comparison to the rotational broadening.

\begin{figure}
   \centering
   \includegraphics[width=9cm]{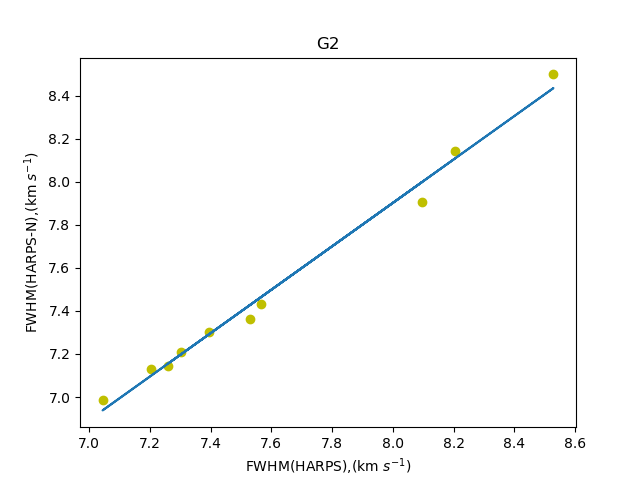}
     \includegraphics[width=9cm]{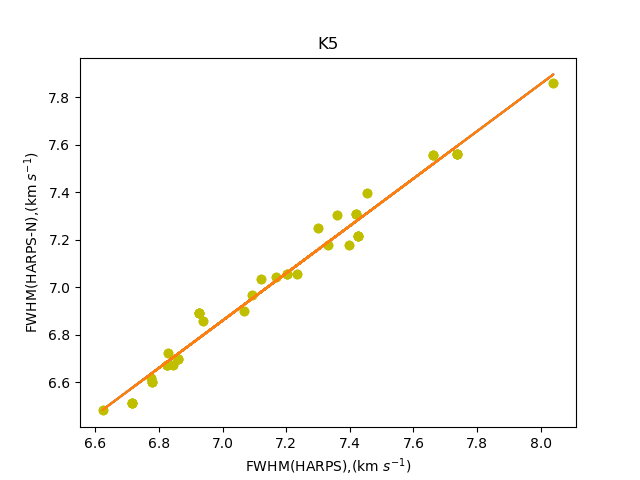}
      \caption{Calibration of FWHM examples in order to derive $\textit{v sin i}$ for common stars sampled in HARPS and HARPS-N data.}
         \label{fwhm}
   \end{figure}

\begin{figure}
   \centering
   \includegraphics[width=9cm]{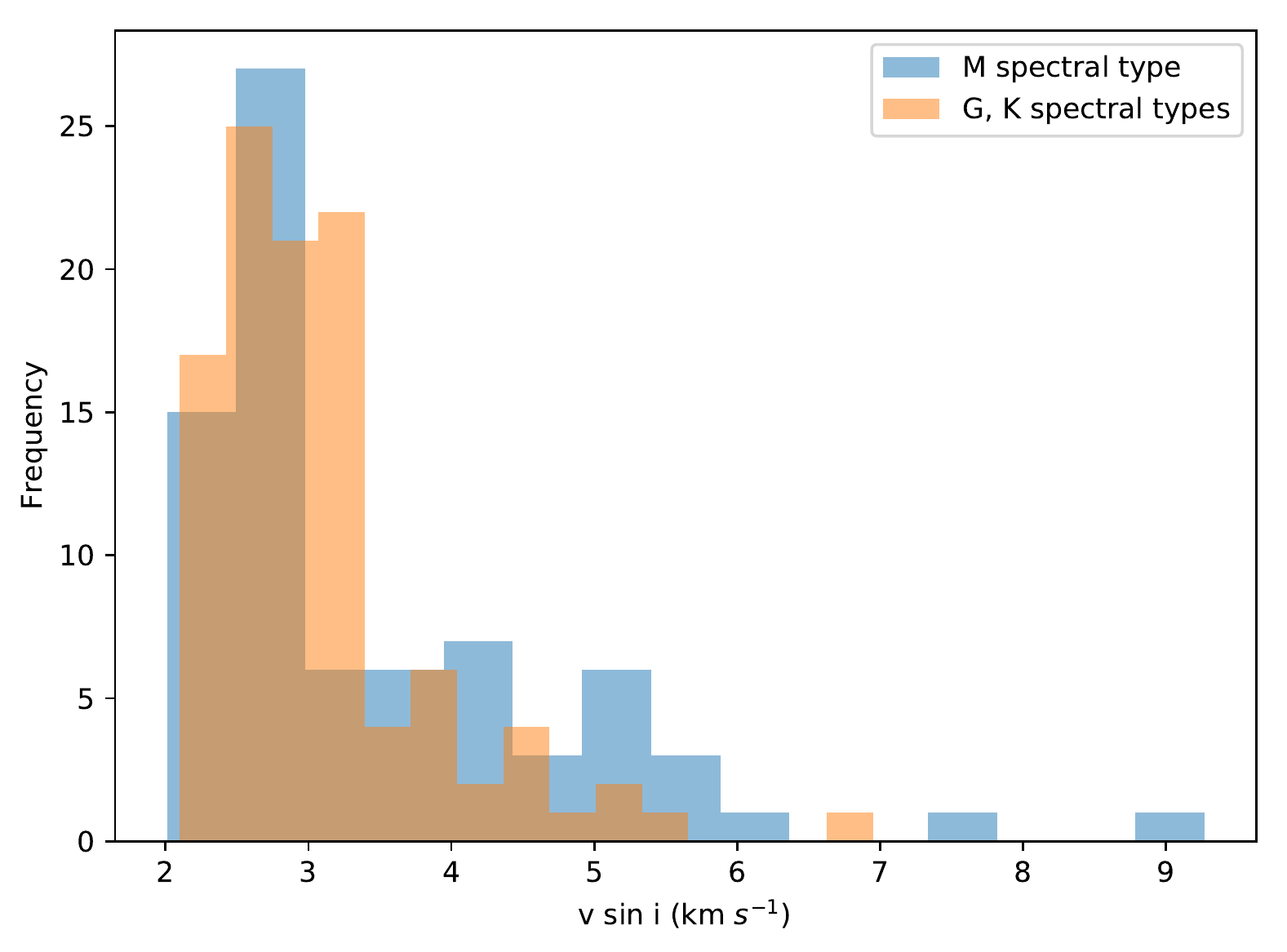}
    
      \caption{Derived $\textit{v sin i}$ distribution for star sample.}
         \label{vsin}
   \end{figure}

We estimated the $\textit{v sin i}$ using the FWHM of the CCF and $B-V$ color and following the procedure described by several studies \citep[e.g.,][with references therein]{2002A&A...392..215S,maldonado2017hades} for HARPS data. We extended the calibration and used the stars that were observed with both HARPS and HARPS-N. Knowing that the instruments are very similar in design and in particular in resolution, a linear relationship was defined between the width of the CCF as measured by the two instruments. For CCF parameters like FWHM, we used the values produced by the previously mentioned standard pipelines of the two instruments. This allowed us to calculate the $v \sin i$ on HARPS-N for each FWHM. This calibration was performed independently for different masks corresponding to different spectral types (see Fig.~\ref{fwhm}). 
We followed the same procedure for UVES spectra as well. For the UVES spectra, we derived the width of the CCF and its associated uncertainty by using the tool provided by \citet{Martins2017}, see Sect.~\ref{data reduction and ccf}. We used a sample of stars in common between HARPS and UVES in order to derive $\textit{v sin i}$ for the stars observed with UVES. \\
In Fig.~\ref{vsin}, we plot the distribution of the $\textit{v sin i}$ measurements. The uncertainties on $\textit{v sin i}$ are derived by error propagation. We considered uncertainties on the CCF width for non-rotating stars and on the observed CCF width \citep{maldonado2017hades}. For stars with more than one observation, we considered the standard deviation of the observed CCF width and the photon noise on the CCF in the error propagation equations. The mean value of the final uncertainties is of 0.47 $km~s^{-1}$, which is compatible with the work of \citet{2010AJ....139..504B} and \citet{maldonado2017hades}. Uncertainties coming from simple error propagation are by construction underestimations and the method needs a deeper analysis on the uncertainties and dependencies on both other astrophysical parameters as well as the systematic errors coming from the simplified model applied to link the $\textit{v sin i}$ to the FWHM of the CCF \citep[][and references therein]{maldonado2017hades,2010AJ....139..504B,2018A&A...612A..49R}. However, it is enough to detect the high rotation velocity stars as it is the aim of the paper. The results for $\textit{v sin i}$ are presented in Table~\ref{result}. We reported values the $\textit{v sin i}$ < 2.0 $km~s^{-1}$ as "< 2.0 $km~s^{-1}$" because of the limited instrumental resolution. The FWHM of a spectrograph with R=100000 corresponds to $\sim$ 2.0 $km~s^{-1}$. As such, one cannot measure rotational velocity values smaller than this value. Most of the targets have low projected rotational velocity except three stars. Twenty-three targets were identified as having a high rotational velocity with $\textit{v sin i}$ > 5.0 $km~s^{-1}$ and were removed from the final sample.
 
\section{Complementary list for the known exoplanet systems \label{known planet}}

On top of the stars discussed above, we decided to give particular attention to the stars with known planets. In particular, we looked for systems that both satisfy our previous criteria and that could host an additional Earth-mass planet in the habitable zone.
We applied the photon noise criteria and visibility on the known G, K, and M spectral type hosting exoplanets. As a source of known exoplanets we used the The Exoplanet Encyclopaedia\footnote{\texttt{$http://www.exoplanet.eu/catalog/$}} and NASA Exoplanet Data Archive\footnote{\texttt{$https://exoplanetarchive.ipac.caltech.edu/index.html$}}. For the stellar parameters we used the SWEET-CAT \citep{santos2013sweet, andreasen2017sweet} to exclude giants ($\log g < 4.1$ $cm~s^{-2}$) and select the stars of G, K, and M spectral types. For each confirmed exoplanet system, we determined the HZ defined by \citet{selsis2007habitable}. Then we investigated the stability of an Earth-mass planet inside HZ using Hill-stability for circular orbits as well as orbits with the given eccentricity in the exoplanets catalogs. For these stars we did not apply activity criteria. We then applied the Eq. 24 from \citet{gladman1993dynamics} as stability criterion
\begin{equation}
       \frac{a_{1} - a_{2}}{a_{1}} > 2.40(\mu_{1}+\mu_{2})^{1/3},
  \end{equation}

  to ten different orbits within the HZ (in regular steps of orbital separation). Here $\mu_{i} = m_{i} / m_{\star}$, $m_{i}$, and $m_{\star}$ are the mass of planets and star, $a_{i}$ is a semi-major axis, and i = 1, 2 for the two planets respectively. For each of them we checked the stability condition for a potential Earth-mass planet in a circular orbit. Our goal was to select and retain only those systems for which at least one of the tested orbits is stable.

On the other hand, for systems with planets in eccentric orbits we used Eq. 21 from \citet{gladman1993dynamics},
\begin{equation}
\left(\mu_{1} + \mu_{2} \frac{a_1}{a_2} \right) \left(\mu_{1} \gamma_{1} + \mu_{2}\gamma_{2} \sqrt{\frac{a_{2}}{a_{1} }} \right)^{2} >\alpha^{3} + 3^{4/3}\mu_{1}\mu_{2}\alpha^{5/3},
  \end{equation}where $\gamma_{i} = \sqrt{1-e^{2}}$, $\alpha = \mu_{1} + \mu_{2}$ and $e_{i}$ is the eccentricity.  
For the systems with more than one planet, each pair of planets was considered. By using the stable-orbits in the HZ of the system, we determined the Earth-mass planet orbital period in the innermost stable orbit in the most optimistic HZ using the generalized Kepler third law:

\begin{equation}
 P/1 yr =  \sqrt{\frac{a^{3}/AU}{M_{\star}/M_{\sun}}}.
  \end{equation}
  
  Then we estimated the semi-amplitude for RV variation (K) using Eq.~\ref{k-formul}. Table~\ref{known-planet-table} lists the parameters for the known exoplanet systems with at least one stable orbit. However using Hill's criterion can be risky because it is valid only for coplanar orbits \citep{gladman1993dynamics} and ignores the possibility of the existence of stable islands due to the secular resonances. Therefore we did not exclude any target from the our list based on this criterion. Further comprehensive analysis of each individual system is required to reach any firm conclusion.

\begin{table*}
\caption{Nineteen known exoplanet systems with free stable orbit in HZ (C=satisfy with circular orbits, EC: satisfy considering eccentric orbits)}             
\label{known-planet-table}      
\centering 
\begin{adjustbox}{width=1\textwidth}
\small         
\begin{tabular}{ccccccccc}     

\hline\hline       
  Star Name &  \teff\ [K] &   $[Fe/H]$  &  $\log g$ [cm$s^{-2}$] &   Spectral type   &   Period HZ [day]  &   K$[cm~s^{-1}]$ &   Stability & $log(R' _{HK}$) and [data sources] \footnotemark[1]  \\
\hline   

HIP43587 	&	5279	$\pm$	62	&	0.33 	$\pm$	0.07 	&	4.37 	$\pm$	0.18	&	G8 	&	130.43 	&	13	&	C	& -5.070  $\pm$ 0.012 [7] \\

HIP98767	&	5584	$\pm$	36	&	0.24 	$\pm$	0.05 	&	4.37 	$\pm$	0.11	&	G8	&	179.46	&	12	&	C & -5.09 [2] 	\\
HIP57443 	&	5629	$\pm$	29	&	-0.29 	$\pm$	0.02 	&	4.44 	$\pm$	0.03	&	G8 	&	198.01 	&	12	&	C & -4.944	 $\pm$ 0.010 [5]	\\
HIP64924 	&	5577	$\pm$	33	&	0.01 	$\pm$	0.05 	&	4.34 	$\pm$	0.11	&	G8 	&	179.06 	&	12	&	C  & -4.920 $\pm$ 0.014 [1] \\
HIP80337 	&	5858	$\pm$	18	&	0.03 	$\pm$	0.01 	&	4.50 	$\pm$	0.03	&	G2 	&	225.18 	&	10	&	C & 4.52 [2]	\\
HIP113357 	&	5804	$\pm$	36	&	0.20 	$\pm$	0.05 	&	4.42 	$\pm$	0.07	&	G5	&	203.02	&	10	&	EC & -4.963 $\pm$ 0.042 [3] 	\\
HIP86796	&	5798	$\pm$	33	&	0.32 	$\pm$	0.04 	&	4.33 	$\pm$	0.08	&	G5	&	209.91	&	10	&	C & -5.000 $\pm$ 0.032 [4]	\\
HIP40693 	&	5402	$\pm$	28	&	-0.06 	$\pm$	0.02 	&	4.40 	$\pm$	0.04	&	K0 	&	158.35 	&	13	&	EC & -4.898 $\pm$ 0.250 [4]	\\
HIP8102 	&	5310	$\pm$	17	&	-0.52 	$\pm$	0.01 	&	4.46 	$\pm$	0.03	&	K0 	&	187.37 	&	13	&	C  & -4.886 $\pm$ 0.202 [8] \\ 
HIP15510	&	5401	$\pm$	17	&	-0.40 	$\pm$	0.01 	&	4.40 	$\pm$	0.03	&	K0 	&	177.95 	&	14	&	EC	& -4.915 $\pm$ 0.016 [9] \\ 
HIP99825 	&	5099	$\pm$	65	&	-0.30 	$\pm$	0.04 	&	4.43 	$\pm$	0.13	&	K0 	&	135.45 	&	14	&	C	& -4.893 $\pm$ 0.060 [10] \\
HIP3093 	&	5182	$\pm$	79	&	0.12 	$\pm$	0.05 	&	4.40 	$\pm$	0.16	&	K0 	&	135.5 	&	14	&	C & -5.02 [2]	\\
HIP16537 	&	5049	$\pm$	48	&	-0.15 	$\pm$	0.03 	&	4.45 	$\pm$	0.09	&	K1 	&	126.01 	&	15	&	C  & -4.440 $\pm$ 0.021 [1]\\
HIP10138	&	5114	$\pm$	61	&	-0.29 	$\pm$	0.04 	&	4.55 	$\pm$	0.13	&	K1 	&	136.18  	&	15	&	EC	& -4.702 $\pm$ 0.120 [11] \\
HIP83043 	&	3519	$\pm$	150	&	-0.04 	$\pm$	0.10 	&	4.76 	$\pm$	0.12	&	M1 	&	27.79  	&	32	&	C & -5.016 [2]	\\
HIP24186	&	3550	$\pm$	50	&	-0.89 	&	--	&	M1	&	33.4	&	46	&	EC & -5.882 $\pm$ 0.051 [6]	\\
HIP85523 	&	3334	$\pm$	110	&	-0.23 	$\pm$	0.09 	&	4.92 	$\pm$	0.21	&	M2	&	31.43	&	41	&	EC & -4.967  $\pm$ 0.110 [6]	\\
HIP106440 	&	3446	$\pm$	110	&	-0.17 	$\pm$	0.09 	&	4.83 	$\pm$	0.15	&	M3 	&	24.24 	&	38	&	EC & -5.323 $\pm$ 0.030 [12]	\\
HIP70890	&	3050	$\pm$	100	&	-0.03 	&	5.0	&	M6 	&	6.54 	&	140	&	C  & -4.963 $\pm$ 0.009 [13] 	\\

\hline    
             
\end{tabular}
\end{adjustbox}
\begin{tablenotes}
\item[1]  [1]: 192.C-0852(A), [2]: NASA Exoplanet Data Archive, [3]: 091.C-0271(A), [4]: 183.C-0972(A), [5]: \citep{da2014long}, [6]: 072.C-0488(E). [7]: 288.C-5010(A), [8]: 093.C-0062(A), [9]: 096.C-0053(A), [10]: 097.C-0021(A), [11]: 60.A-9700(G), [12]: 198.C-0838(A), [13]: 191.C-0505(A)
 \end{tablenotes}

\end{table*}

\section{Stellar characterization and chemical abundance \label{stellarparam}}

Stellar parameters are not only crucial for accurate characterization of exoplanets but there is also increasing evidence that the characteristics of exoplanets are correlated to several host stars' properties \citep[e.g.,][]{adibekyan2016type,mayor2004coralie,fischer2005planet,reffert2015precise}. In recent studies, chemical abundance has provided useful constraints for planet structure and composition, and to investigate planet formation theories \citep[e.g.,][]{santos2017constraining}. For this aim, we derived stellar parameters (\teff: effective temperature, [Fe/H]: metallicity, $\log g$: surface gravity, and $\xi_{\mathrm{t}}$: microturbulence) and chemical abundances for all of the stars in the sample.\\

The stellar parameters and chemical abundances for G and K dwarfs are determined with the procedure described in \citet[e.g.][]{Sousa-14, Adibekyan-12c}. In short,  we first automatically measured the equivalent widths (EWs) of the spectral lines using the Automatic Routine for line Equivalent widths in stellar Spectra, ARES v2 code \citep{Sousa-15} \footnote{The last version of ARES code (ARES v2) can be downloaded at http://www.astro.up.pt/$\sim$sousasag/ares}. For the more problematic elements we performed careful visual inspection of the determination of EWs.
Then the spectroscopic parameters and chemical abundances were derived using the classical curve-of-growth analysis method forcing excitation and ionization equilibrium under assumption of local thermodynamic equilibrium (LTE). 
We used the grid of the ATLAS9 plane-parallel model of atmospheres \citep{Kurucz-93} and the 2014 version of MOOG\footnote{The source code of MOOG can be downloaded at
http://www.as.utexas.edu/$\sim$chris/moog.html} radiative transfer code \citep{Sneden-73}. The uncertainties of the stellar parameter and chemical abundances are also derived as in the references above. It is important to note that they represent the internal precision of  the technique and not  the absolute accuracies that can be significantly worse due to several factors as discussed in \citet[e.g.][]{2016ApJS..226....4H,2017A&A...601A..38J}.

Abundances of the volatile elements, C and O, were only derived for a subsample of the stars and following the method of \cite{Delgado-10, Bertrandelis-15}. As discussed in those works the derivation of abundances for these elements with the EW method and the lines employed here is only possible for unevolved stars with \teff $\gtrsim$ 5100 K. For Carbon, we measured the EWs of two atomic lines with ARES v2 code but under a visual inspection since the line at 5052 \AA{} gets weaker as \teff\ decreases. Given that the very few oxygen lines may be strongly blended, we choose to manually measure them with the task \textit{splot} in Image Reduction and Analysis Facility
, IRAF. The line at 6158 \AA{} gets very weak for cooler stars and the forbidden line at 6300 \AA{} is blended with Ni and CN lines (which become stronger as \teff\ decreases). Then the abundances were derived using the 2014 version of MOOG as they were for the other elements.

\begin{figure}
\begin{center}
\begin{tabular}{c}
\includegraphics[angle=0,width=1.0\linewidth]{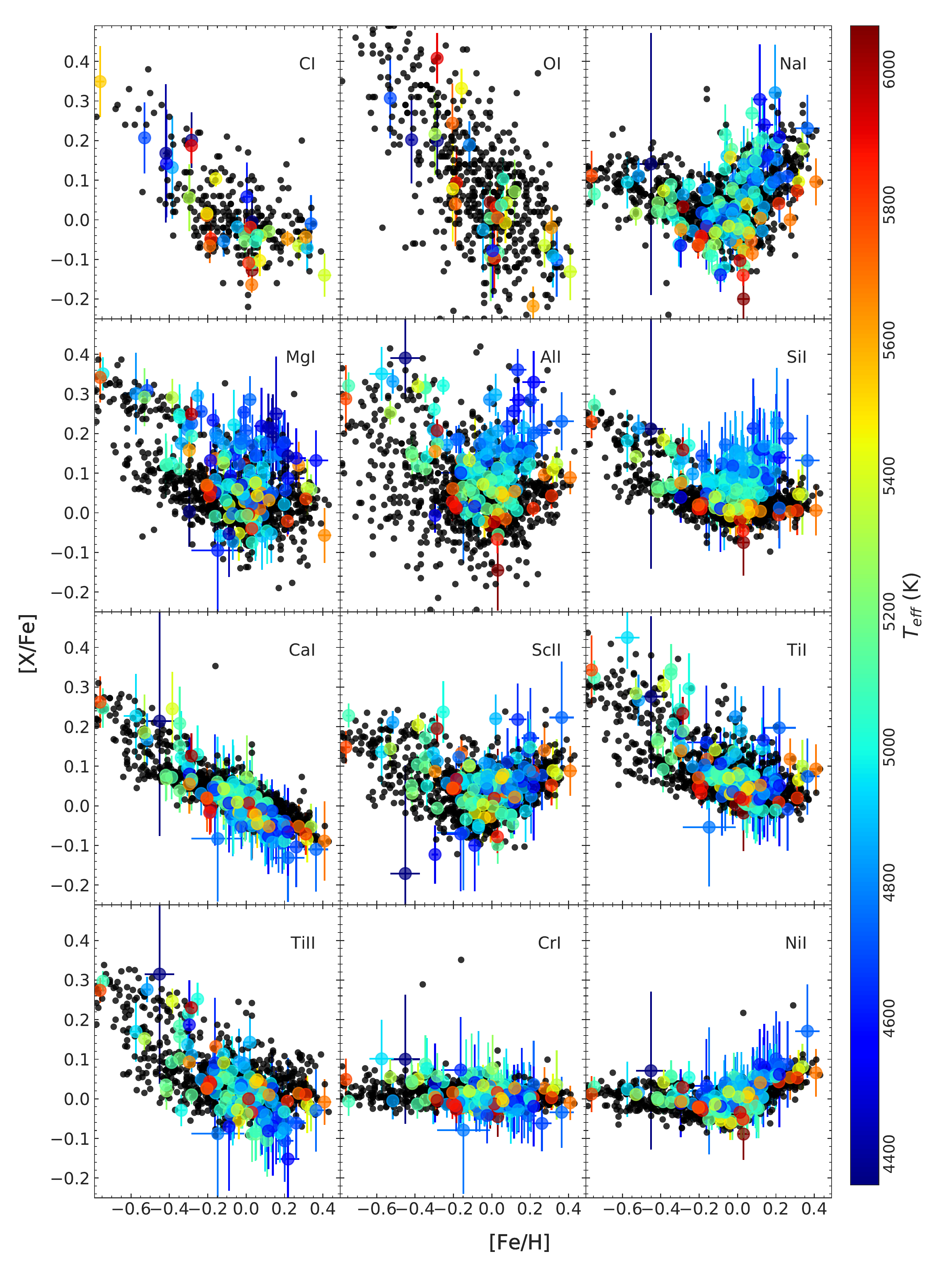}
\end{tabular}
\end{center}
\vspace{-0.9cm}
\caption{Abundance ratio [X/Fe] against stellar metallicity for current sample (color-coded by  \teff) and for HARPS-sampled field stars (black).}
\label{plot_elfe_feh_teff}
\end{figure}

\begin{figure}
\begin{center}
\begin{tabular}{c}
\includegraphics[angle=0,width=1.0\linewidth]{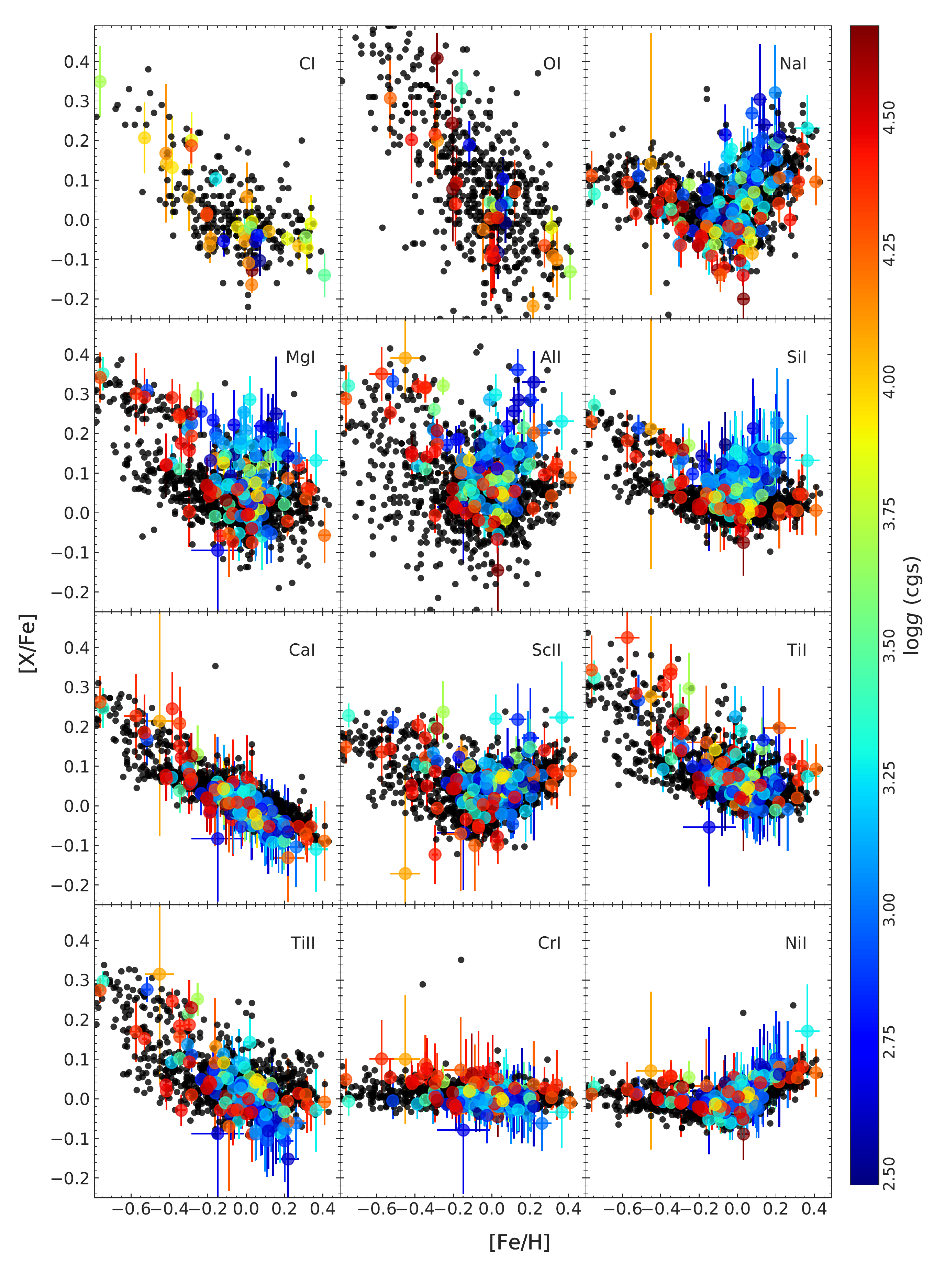}
\end{tabular}
\end{center}
\vspace{-0.9cm}
\caption{Abundance ratio [X/Fe] against stellar effective temperature for current sample (color-coded by $\log g$) and for HARPS-sampled field stars (black).}
\label{plot_elfe_feh_logg}
\end{figure}

In Figs.~\ref{plot_elfe_feh_teff} and ~\ref{plot_elfe_feh_logg} we present dependence of the [X/Fe] abundance ratios on the stellar metallicity. The comparison is done for the large primary sample including stars with $\log g$ < 4.1 $cm~s^{-2}$. For comparison, the HARPS sample 
of FGK dwarf stars from \citet{Adibekyan-12c} is plotted as well. The abundances of atomic Carbon and Oxygen for the HARPS sample stars were derived in \citet{Delgado-10} and \citet{Bertrandelis-15}, 
respectively\footnote{Note that \citet{Suarez-17} also derived Carbon abundance for the HARPS sample stars from the CH band at 4300 \AA{}.}. Figure~\ref{plot_elfe_feh_teff} shows
that the two samples generally follow the same trend set by the Galactic chemical evolution. However, some discrepancies are apparent and worth mentioning. It seems that at metallicities above solar value
Al and especially Si abundances relative to iron are higher for the current sample when compared to the HARPS sample stars. For the same metallicities, Ca and TiII seem to be under-abundant when compared 
to their HARPS counterparts. Most probably this is because many of the stars of the first sample are giant, evolved stars. Figure~\ref{plot_elfe_feh_logg} shows that indeed, most of the Al- and Si-enhanced,
and Ca- and TiII-poor stars are evolved stars with $\log g < 3.5$ dex. Several studies observed such differences in chemical abundances between main-sequence dwarfs and evolved stars 
\citep[e.g.,][]{Friel-03, Villanova-09, Adibekyan-15}. These differences might have  astrophysical origins \citep[e.g.,][as in the case of Sodium]{Tautvaisien-05}, 
they might be related to the spectroscopic analysis methods and use line-lists \citep[e.g.,][]{Santos-09} and/or non-LTE effects, which are stronger for evolved stars \citep[e.g.,][]{Bergemann-13}.
Some of the elements also show a dependence on effective temperature that has already been noticed and discussed in the literature
\citep[e.g.][]{Gilli-06, Lai-08, Adibekyan-12c}. These trends may be related to the stronger line blendings at low temperatures, deviations from excitation
and/or ionization equilibrium, or an incorrect T-$\tau$ relationship in the adopted model atmospheres \citep[see][for more details and references]{Adibekyan-12c}

Figure~\ref{plot_elfe_feh_teff} shows that the most iron-poor stars are enhanced in $\alpha$ elements and probably belong to the Galactic thick disk \citep[e.g.,][]{Adibekyan-13, Recio-Blanco-14}.
This is very interesting to note since it was found that planets orbiting iron-poor stars prefer the $\alpha$-enhanced ones \citep[e.g.,][]{Haywood-09, Adibekyan-12a, Adibekyan-12b}. Enhancement 
of the $\alpha$ elements, such as Mg and Si, relative to iron can play a very important role not only for the formation of planets, but also on the composition of the formed planets \citep{Santos-17}.\\
For the M dwarfs for which spectra were available, we used a modified version of the method and software developed by \citet{neves2014metallicity} to \teff\ and [Fe/H]\footnote{\texttt{http://www.astro.up.pt/resources/mcal/Site/MCAL.html}}. The \teff\ scale was updated and its initial values were calculated as the average of the ($V-J$), ($V-H$), and ($V-K$) calibrations taken from \citet{2012ApJ...757..112B} . The corresponding errors were reported using Tables 5 and 6 in \citet{neves2014metallicity}. \ 
The values are presented in Tables \ref{stellar-param-table},  \ref{chemichal1},  and \ref{chemichal2}. We presented $\log g$ using Eq. \ref{logg_hip} and \teff using color calibrations where the sufficient spectra were not available and the targets were already discarded by the other criteria from the final list. The error bar for these targets were not calculated.

 \subsection{Composition of planetary building blocks}
  
\begin{figure}
\begin{center}
\begin{tabular}{c}
\includegraphics[angle=0,width=1\linewidth]{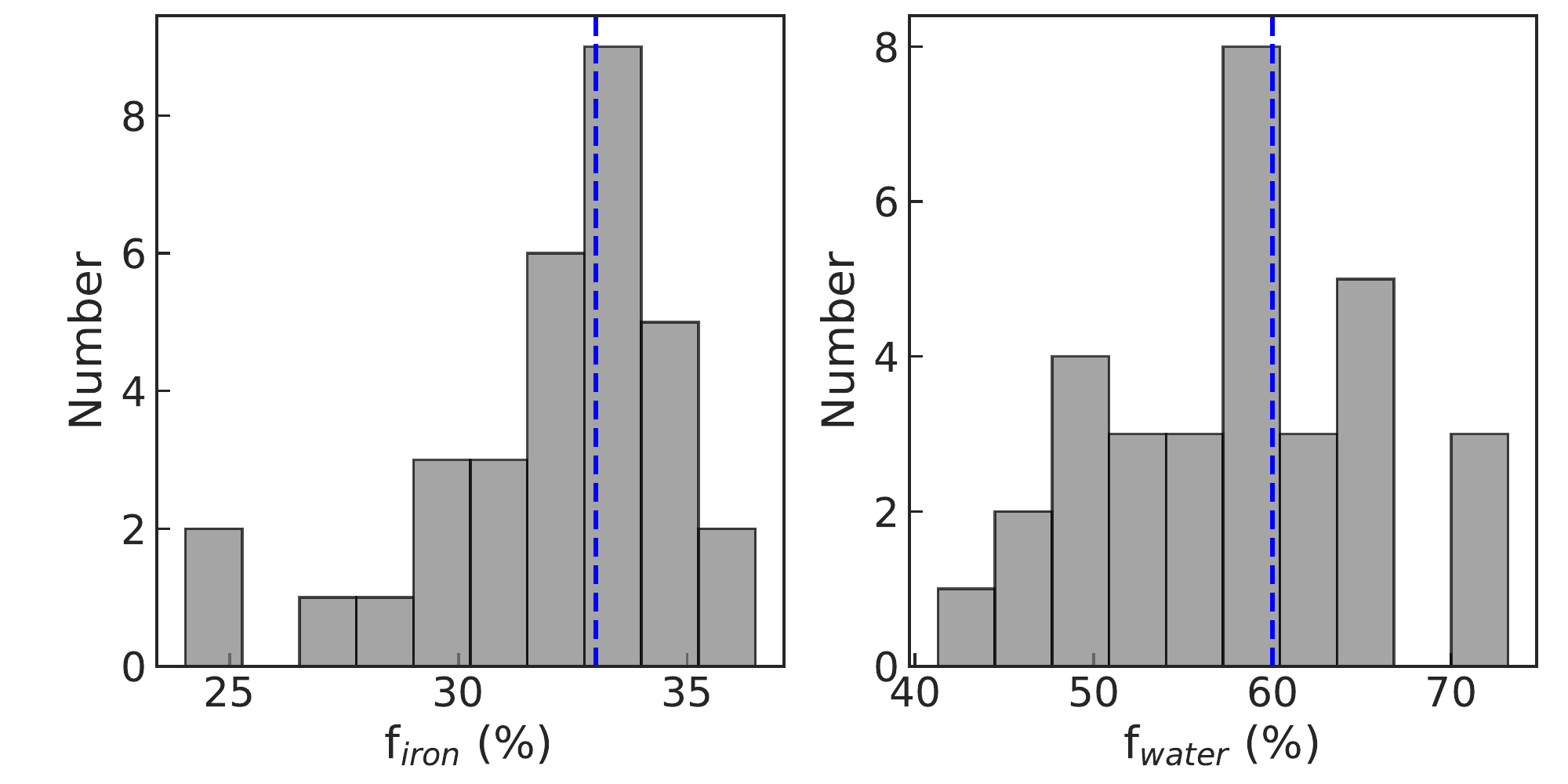}
\end{tabular}
\end{center}
\vspace{-0.3cm}
\caption{Expected iron and water mass fraction distributions of planetary building grains around sample stars. The values of these quantities derived 
for the solar system with our model are shown with dashed blue line.}
\label{plot_f_wf}
\end{figure}

 For a sample of 32 solar-type stars we derived abundances of the main rock-forming refractory (Mg, Si, and Fe) and volatile (O, C) elements. We used the model described in
  \citet{Santos-17} to compute the expected iron-to-silicate mass fraction (f$_{iron}$) and the water mass fraction (\textit{wf}) in the planetary building blocks. The derived values 
  of these quantities are presented in Table ~\ref{tab:f_wf} and their distributions are shown in Fig.~\ref{plot_f_wf}. The distribution of f$_{iron}$ and \textit{wf}
  are very similar to the distribution of the same quantities derived by \citet{Santos-17} for the Galactic thin disk stars in the solar neighborhood.
  
\begin{table}[t!]
\caption{\label{tab:f_wf} Iron-to-silicate and water mass fraction of planet building blocks. ($\star$ : host confirmed exoplanet)}
\begin{tabular}{ccc}
\hline
      star &   $f_{iron}~[\%]$ &   $\textit{wf}~[\%]$    \\
\hline
\hline
HIP13402 & 35.0 $\pm$ 3.1 & 58.1 $\pm$ 7.1 \\
HIP16852 & 31.0 $\pm$ 2.0 & 57.3 $\pm$ 4.6 \\
HIP22263 & 36.5 $\pm$ 1.3 & 59.6 $\pm$ 7.7 \\
HIP23835 & 29.0 $\pm$ 1.2 & 73.1 $\pm$ 2.6 \\
HIP43587$\star$ & 31.6 $\pm$ 3.0 & 46.5 $\pm$ 8.0 \\
HIP64924$\star$ & 33.3 $\pm$ 1.8 & 49.3 $\pm$ 8.8 \\
HIP77257 & 33.0 $\pm$ 1.1 & 58.5 $\pm$ 2.8 \\
HIP83541 & 31.5 $\pm$ 2.6 & 47.4 $\pm$ 6.5 \\
HIP95447 & 35.3 $\pm$ 2.8 & 48.7 $\pm$ 5.5 \\
HIP98066 & 34.7 $\pm$ 1.6 & 51.9 $\pm$ 5.1 \\
HIP98767$\star$ & 31.2 $\pm$ 2.1 & 49.4 $\pm$ 4.5 \\
HIP101345 & 32.9 $\pm$ 1.0 & 59.3 $\pm$ 3.7 \\
HIP101916 & 33.5 $\pm$ 1.9 & 56.2 $\pm$ 4.0 \\
HIP113357$\star$ & 34.8 $\pm$ 2.0 & 41.3 $\pm$ 4.7 \\
HIP37606 & 31.9 $\pm$ 1.4 & 57.8 $\pm$ 4.2 \\
HIP107649 & 34.8 $\pm$ 1.9 & 50.8 $\pm$ 6.5 \\
HIP86796$\star$ & 33.2 $\pm$ 2.0 & 55.1 $\pm$ 4.0 \\
HIP114699$\star$ & 30.2 $\pm$ 1.1 & 66.1 $\pm$ 3.6 \\
HIP29271 & 33.4 $\pm$ 1.8 & 61.4 $\pm$ 5.6 \\
HIP1599 & 32.2 $\pm$ 1.4 & 61.7 $\pm$ 6.1 \\
HIP8102$\star$ & 24.0 $\pm$ 1.8 & 66.6 $\pm$ 6.5 \\
HIP10138$\star$ & 27.6 $\pm$ 1.7 & 64.0 $\pm$ 6.0 \\
HIP15330 & 32.0 $\pm$ 1.9 & 60.1 $\pm$ 9.1 \\
HIP15371 & 32.4 $\pm$ 1.6 & 70.8 $\pm$ 5.6 \\
HIP38908 & 24.6 $\pm$ 1.3 & 73.2 $\pm$ 3.4 \\
HIP40693$\star$ & 32.6 $\pm$ 1.6 & 53.8 $\pm$ 8.1 \\
HIP56452 & 29.7 $\pm$ 2.0 & 66.0 $\pm$ 6.9 \\
HIP57443 & 28.6 $\pm$ 1.6 & 65.8 $\pm$ 6.6 \\
HIP81300 & 34.1 $\pm$ 2.3 & 51.2 $\pm$ 9.2 \\
HIP110109 & 33.0 $\pm$ 1.2 & 58.7 $\pm$ 7.4 \\
\hline
\end{tabular}
\end{table}

\section{Stellar rotational period estimation for M dwarfs  \label{period}}

The rotation period of M dwarfs can be of the same order of magnitude as the period of a planet in the HZ. The presence of active regions on M dwarfs can thus produce an RV signal that can mimic the signal produced by an Earth-mass exoplanet inside the HZ or at any harmonic of the rotation period \citep[e.g.,][]{boisse2009stellar}.

We derived the expected rotational period of M dwarfs using activity-rotation relationships as a first estimation. Several studies show a significant correlation between the activity and rotation for M dwarfs \citep[e.g.,][]{mascareno2018hades,astudillo2017magnetic,mascareno2016magnetic}. We used  Eq.1 and Table 1 from \citet{suarez2015rotation} for the rotational period determination from the $log(R' _{HK})$. In Table~\ref{period-M}, we list the rotational period estimated in this way, as well as the orbital period corresponding to the inner boundary of the HZ for our sample of M dwarfs. We derived the error using the standard error propagation of the input parameters' errors. For low-activity stars ($log(R' _{HK})$ < -4.80), the rotation periods shows a large dispersion, specially for the early M spectral types. Therefore, in order to take into account the actual dispersion of the data points in Fig. 12 from \citet{mascareno2018hades}, we considered an uncertainty in the rotation period inflated by a factor of three.

 We are aware that the rotation periods obtained in this way are a first-order estimation. Although the statistical strength of the relationship for the purpose of this study is sufficient enough, we did not exclude any target from the final list based on this criterion. There are some stars for which extreme caution is necessary because the rotational period of the stars are very close to the HZ inner edge's orbital period. For these targets distinguishing the planetary signal from RV variations of stars due to its rotation will be quite challenging. The measurements can be improved in the future studies using either photometric or spectroscopic methods.

\begin{table*}
\caption{Sample of M dwarf in final list with rotational periods and HZ parameters.}        
\label{period-M}      
\centering       
\small   
\begin{tabular}{ccccc}     
\hline\hline       
  Star Name &  $P_{rot}$ [day] & Stellar mass [$M_{\sun}$] and Reference  &  $P_{inner}$[day]  & HZ [AU] \\
\hline     
HIP439	&	79	$\pm$	6	&	0.39$\pm$0.03 [1]	&	25$\pm$1	&	[ 0.12 , 0.32 ]\\
HIP1242	&	192	$\pm$	8	&	0.14 [2]	&	12	&	[ 0.05 , 0.14 ]\\
HIP22627	&	64	$\pm$	6	&	0.37$\pm$0.06 [3]	&	20$\pm$2	&	[ 0.1 , 0.28 ]\\
HIP23708	&	33	$\pm$	5	&	0.59$\pm$0.02 [4]	&	55$\pm$1	&	[ 0.24 , 0.61 ]\\
HIP29295	&	29	$\pm$	5	&	0.58$\pm$0.06 [1]	&	41$\pm$2	&	[ 0.19 , 0.5 ]\\
HIP40239	&	161	$\pm$	8	&	0.61 [2]	&	49	&	[ 0.22 , 0.57 ]\\
HIP42748	&	81	$\pm$	6	&	0.45$\pm$0.05 [4]	&	60$\pm$3	&	[ 0.23 , 0.59 ]\\
HIP45908	&	31	$\pm$	5	&	0.55$\pm$0.03 [1]	&	38$\pm$1	&	[ 0.18 , 0.46 ]\\
HIP51317	&	46	$\pm$	5	&	0.44$\pm$0.03 [1]	&	26$\pm$1	&	[ 0.13 , 0.34 ]\\
HIP53020	&	88	$\pm$	6	&	0.26$\pm$0.02 [1]	&	14$\pm$1	&	[ 0.07 , 0.2 ]\\
HIP62452	&	104	$\pm$	7	&	0.32$\pm$0.02 [1]	&	16$\pm$1	&	[ 0.09 , 0.23 ]\\
HIP65859	&	30	$\pm$	5	&	0.53$\pm$0.03 [1]	&	34$\pm$1	&	[ 0.17 , 0.43 ]\\
HIP67164	&	80	$\pm$	6	&	0.42$\pm$0.06 [5]	&	14$\pm$1	&	[ 0.09 , 0.23 ]\\
HIP71253	&	85	$\pm$	6	&	0.28$\pm$0.02 [1]	&	15$\pm$1	&	[ 0.08 , 0.21 ]\\
HIP85523	&	38	$\pm$	5	&	0.35$\pm$0.03 [1]	&	21$\pm$1	&	[ 0.1 , 0.27 ]\\
HIP86214	&	35	$\pm$	5	&	0.27$\pm$0.02 [1]	&	15$\pm$1	&	[ 0.08 , 0.21 ]\\
HIP86287	&	73	$\pm$	6	&	0.45$\pm$0.03 [1]	&	28$\pm$1	&	[ 0.14 , 0.36 ]\\
HIP87937	&	73	$\pm$	6	&	0.16$\pm$0.01 [1]	&	9$\pm$1	&	[ 0.04 , 0.12 ]\\
HIP88574	&	46	$\pm$	5	&	0.48$\pm$0.03 [1]	&	30$\pm$1	&	[ 0.15 , 0.38 ]\\
HIP92403	&	30	$\pm$	5	&	0.17$\pm$0.01 [1]	&	11$\pm$1	&	[ 0.05 , 0.14 ]\\
HIP93069	&	49	$\pm$	5	&	0.53 [2]	&	59	&	[ 0.24 , 0.62 ]\\
HIP93101	&	28	$\pm$	4	&	0.43$\pm$0.05 [4]	&	51$\pm$3	&	[ 0.2 , 0.52 ]\\
HIP99701	&	26	$\pm$	4	&	0.59 [2]	&	42	&	[ 0.2 , 0.51 ]\\
HIP106440	&	52	$\pm$	5	&	0.45$\pm$0.03 [1]	&	28$\pm$1	&	[ 0.14 , 0.35 ]\\
HIP113020	&	112	$\pm$	7	&	0.34$\pm$0.02 [1]	&	18$\pm$1	&	[ 0.09 , 0.25 ]\\
HIP115332	&	44	$\pm$	5	&	0.4$\pm$0.02 [1]	&	20$\pm$1	&	[ 0.1 , 0.28 ]\\
HIP67155	&	38	$\pm$	5	&	0.5$\pm$0.03 [1]	&	29$\pm$1	&	[ 0.14 , 0.37 ]\\
\hline                  
\end{tabular}
\begin{tablenotes}
\item[1]  [1]: \cite{neves2013metallicity}, [2]: \cite{gaidos2014trumpeting}, [3]: \cite{mann2015constrain}, [4]: \cite{mints2017unified}, [5]: \cite{gaidos2014m}
 \end{tablenotes} 
\end{table*}

\section{Summary and conclusion \label{sum}}

\begin{figure}
   \centering
   \includegraphics[width=9.8cm]{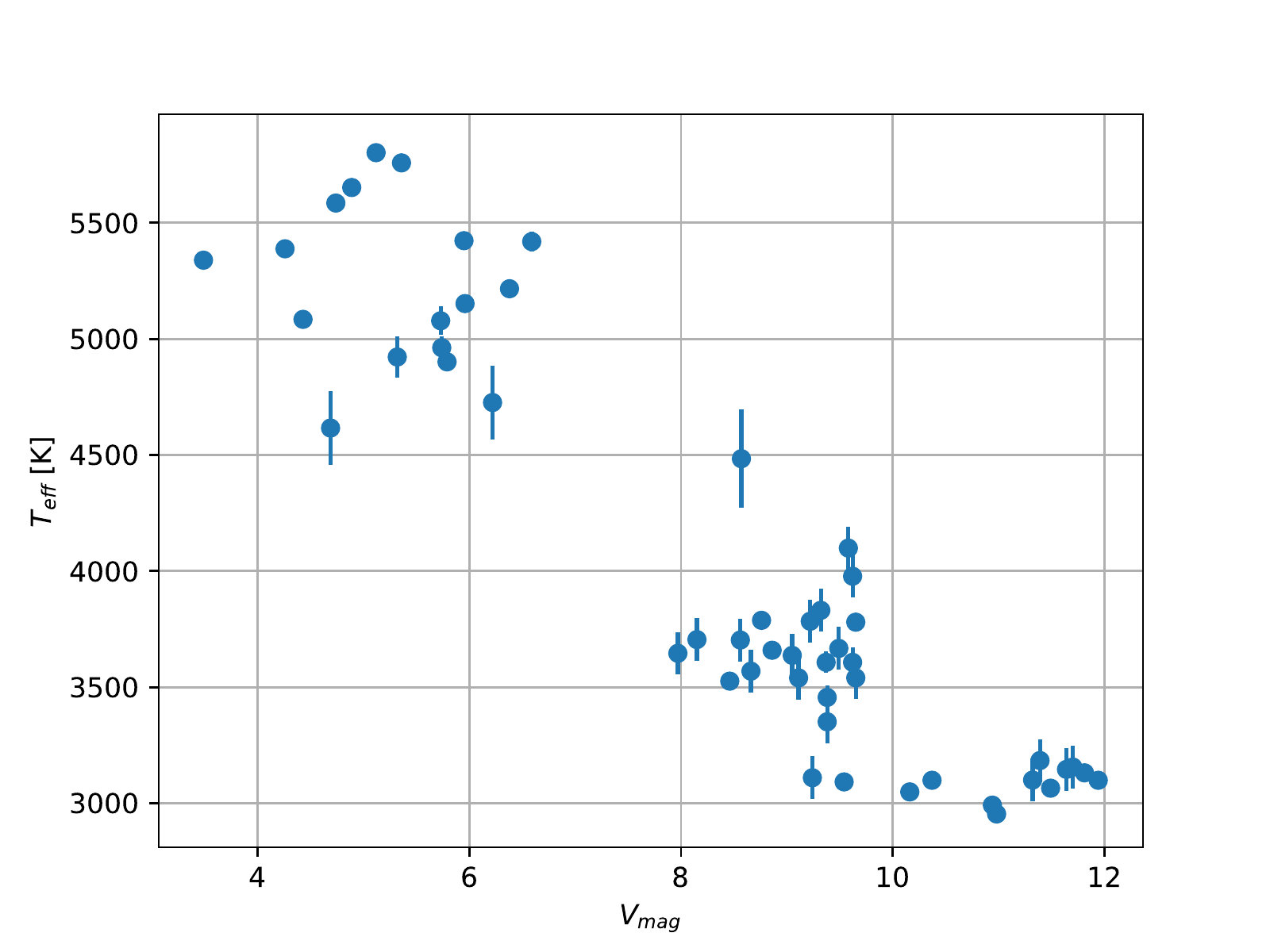}
   
      \caption{Visual magnitude vs \teff\ [K] for final selected sample of stars}
         \label{final-v-m}
   \end{figure}

\begin{figure}
   \centering
   \includegraphics[width=9cm]{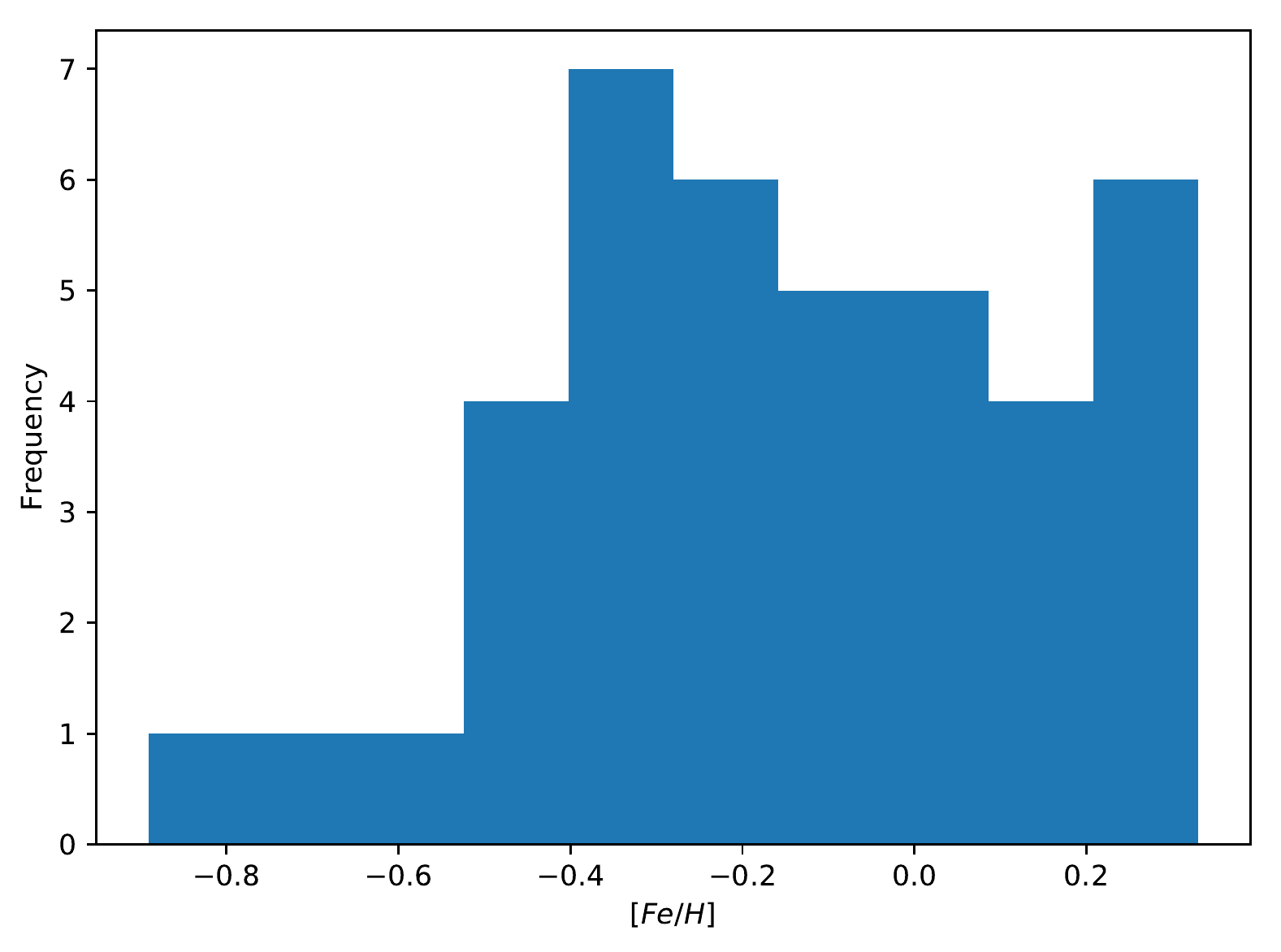}
   
      \caption{[Fe/H] distribution for final selected sample of stars where metallically values are available.}
         \label{final-feh}
   \end{figure}

We present a detailed spectroscopical characterization of a sample of bright stars that are suitable targets for a precise RV planet search program focused on Earth-mass planets in their hosts habitable zone using ESPRESSO. To build this sample, we used new and archival high-resolution spectroscopic data. We screened a large sample of G, K, and M stars for visibility, spectroscopic contamination, chromosphere activity using $log(R'_{HK})$, projected rotational velocity using $\textit{v sin i}$, stellar parameters, and chemical abundances. After considering previous criteria, for the known host planetary systems we investigated the availability of a free, stable orbit for an Earth-mass planet inside the HZ.\\
Twenty-seven stars were discarded according to the signature of stellar companions in CCF criteria. Forty-three stars with high activity $log(R'_{HK})$ > -4.80 were excluded from the list due to the expected large activity-induced RV variations. We identified twenty-three stars with $\textit{v sin i}$ > 5.0 $km~s^{-1}$ and excluded them from the final list. The list of the 45 best targets is presented in Table~\ref{sellected-stars}. For the known host planetary systems which were described in Sect.~\ref{known planet}, nineteen best candidates are presented in Table~\ref{known-planet-table}. Five G, fifteen K, and twenty-seven M spectral type stars have $V$ between 3.49 and 11.94. Figures~\ref{final-v-m} and \ref{final-feh} present the distribution of [Fe/H] for forty stars with available chemical abundance values. \\

When performing our study we realized that many bright, main-sequence GKM stars did not have sufficiently high $S/N$ spectra in the ESO archive to allow spectroscopic analysis. The spectra obtained by us were made available immediately, and the reduced spectra, made available now, can be of interest to a wide range of studies. All the spectra are available for download from the ESO archive and also from our Porto data server for the case of HARPS-N observations\footnote{\texttt{www.astro.up.pt/$\sim$saeed/HARPS$-$N/HARPS$-$N.zip}}.\\
Concerning the final target list, applying the criteria on spectroscopic contamination, stellar activity, and chromospheric activity levels, and discarding evolved stars by a direct or indirect assessment of the log$g$, led us to discard a large fraction of the stars. At the end of the selection steps, we retained only $\sim$45 of the $\sim$249 stars originally surveyed spectroscopically. This selection, which was devised to be independent of spectral type in any other aspect than photon noise, let to a catalog that is composed of roughly of 60\% of  M dwarfs. This is the first study that leads to the conclusion that, even when trying to be as independent as possible of spectral-type biases, when focusing on objective criteria such as photon noise and signal detectability from first principle, M dwarfs emerge as optimal RV targets, even at optical wavelengths. \\

Most of the stars (27) in the final list are of M spectral type. M dwarfs are good candidates not only for gravitational RV detection because of their relativity low mass, but also for having a closer HZ. Hypothetical planets around M dwarfs would have relatively shorter orbital periods, making it is easier to cover their orbital period. However M dwarfs with rotational periods close to the orbital period of an Earth-class planet in their HZ will be considered as of lower priority for observations. The sample of M dwarfs in our final list consists mostly of early-type and late-type spectral sub-types with very few in the intermediate sub-types. M dwarfs are not a homogeneous class of stars and there are many intermediate M dwarfs in this magnitude regime \citep[see Fig. 5 in][]{2016A&A...586A.101F}, so this came probably from statistical fluctuations arising from small number statistics. It should be mentioned that the temperature of the M dwarfs scatter over a few hundreds of kelvins in the different catalogs therefore we cannot compare the parameters in the the sub-spectral M dwarfs. The effective temperature determination of M dwarfs is affected by an uncertainty at the level of 100 K. As such, a detailed  spectral type assignment based on effective temperatures is impossible, and we warn the reader against an over-interpretation of the data.\\

We considered stars for which the photon noise allowed the detection of an Earth-class planet inside the HZ. In the absence of sources of error other than the photon noise, an Earth-class planet is detectable in almost all of the candidates with orbital periods of $\approx$ 6 to $\approx$ 223 days by assuming the expected RV precision of $\rm 10 \;cm \;s^{-1}$. Eighty percent  of the ESPRESSO GTO is dedicated to exoplanet searches and one-third of this time is allocated to the blind RV survey (65 nights). By considering 9 hours as an average of night duration, an average of 100 data points in order to cover the orbital periods and 15-minute exposure time, we can intensively observe around 23 targets in this survey. \\
The composition of the planetary building block analysis suggests that the potential terrestrial planets orbiting the majority of these stars will have an iron core-size similar to that of the Earth, but will have a lower fraction of water content. It is also interesting to note that (based only on the chemistry of the host stars) terrestrial planets with very different compositions (e.g., core size) and water content when compared to the solar system, are also expected to be detected around some of the targets. Here we should note that the f$_{iron}$ and \textit{wf} derived with this model might not necessarily be exactly the same as the water content and core size of planets around the target stars \citep[see][]{Santos-17}. These quantities are only indicative of the content of water in the protoplanetary disk and the iron-to-silicate ratio that might be translated into a core-to-mantle ratio. \\
Presently there is no clear indication if the presence of close-in planets (that have been detected e.g., with HARPS) is indicative of the presence of low-mass planets in the habitable zone. On one side, we know from Kepler data that some planetary systems are compact and dynamically full \citep{2014ApJ...790..146F}. If this is a general feature, then the presence of close-in planets could be a positive indication of the presence of planets further out (whatever their mass). Since a giant planet in the HZ of such systems would already have been detected with HARPS, planets in the HZ should be rather small. On the other side, we know from the only confirmed Earth-like planet in the HZ (the Earth itself), that the absence of close-in planets is compatible with the presence of a low-mass planet in the HZ. Finally there are indications, from planet formation models, that the absence of close-in planets could be positively correlated with the presence of low-mass planets in the HZ (Alibert et al., in prep). Since there is no clear consensus on the correlation between the presence of already detected planets and the probability of having a planet in the HZ, we decided to  include both types of stars (with or without already known planets) in our target list.

 All derived parameters for the whole sample are presented in Table~\ref{result} for our new observation and archives respectively. We derived the stellar parameters and chemical abundances for this sample of stars (Tables~\ref{stellar-param-table}, \ref{chemichal1}, and \ref{chemichal2}) where it was possible.

\begin{table*}
\caption{Forty-five best final targets for RV observation}             
\label{sellected-stars}      
\centering          
\begin{tabular}{cccccccc}     
\hline\hline       
Name & R.A [hh:mm:ss] & Dec [$^{\circ}$:$^\prime$:$''$]  & $\textit{v sin i}$ [km $s^{-1}$]  &   $log(R' _{HK})$  &  $\log I_{H\alpha _{index}}$ & $V $  & Sp Type\\ 
\hline

HIP86796	&	17 44 08.7	&	-51 50 00.9	&	2.95 $\pm$ 0.90	&	-5.15	$\pm$	0.01	&	-1.37	$\pm$	0.07	&	5.12	&	G5\\
HIP83541	&	17 04 27.8	&	-28 34 55.3	&	< 2.0	&	-5.02	$\pm$	0.01	&	-1.43	$\pm$	0.01	&	6.59	&	G8 \\
HIP57443	&	11 46 32.2	&	-40 30 04.8	&	< 2.0	&	-4.88	$\pm$	0.01	&	-1.49	$\pm$	0.06	&	4.89	&	G8 \\
HIP64924	&	13 18 25.0	&	-18 18 31.0	&	< 2.0	&	-4.91	$\pm$	0.01	&	-1.45	$\pm$	0.01	&	4.74	&	G8\\
HIP40693	&	08 18 23.8	&	-12 37 47.2	&	< 2.0	&	-4.94	$\pm$	0.01	&	-1.46	$\pm$	0.09	&	5.95	&	G8\\
HIP41926	&	08 32 52.2	&	-31 30 09.7	&	--	&	-4.91	$\pm$	0.01	&	-1.43	$\pm$	0.10	&	6.38	&	K0 \\
HIP15510	&	03 19 53.2	&	-43 04 17.6	&	--	&	-5.00	$\pm$	0.01	&	-1.45	$\pm$	0.01	&	4.26	&	K0 \\
HIP8102	&	01 44 05.1	&	-15 56 22.4	&	--	&	-4.90	$\pm$	0.01	&	-1.44	$\pm$	0.07	&	3.49	&	K0 \\
HIP19849	&	04 15 17.6	&	-07 38 40.4	&	< 2.0	&	-4.92	$\pm$	0.01	&	-1.41	$\pm$	0.01	&	4.43	&	K1 \\
HIP99825	&	20 15 16.6	&	-27 01 57.1	&	< 2.0	&	-4.96	$\pm$	0.01	&	-1.41	$\pm$	0.01	&	5.73	&	K1 \\
HIP56452	&	11 34 30.0	&	-32 50 00.0	&	--	&	-4.83	$\pm$	0.04	&	-1.44	$\pm$	0.10	&	5.96	&	K1\\
HIP3765	&	00 48 22.5	&	+05 17 00.2	&	--	&	-4.93	$\pm$	0.01	&	-1.41	$\pm$	0.01	&	5.74	&	K2 \\
HIP99461	&	20 11 11.6	&	-36 05 50.6	&	< 2.0	&	1.50			&	-1.41	$\pm$	0.02	&	5.32	&	K2\\
HIP12114	&	02 36 03.8	&	+06 53 00.1	&	2.00 $\pm$ 0.27	&	-4.96	$\pm$	0.01	&	-1.38	$\pm$	0.04	&	5.79	&	K2\\
HIP23311	&	05 00 48.7	&	-05 45 03.5	&	< 2.0	&	-5.00	$\pm$	0.01	&	-1.39	$\pm$	0.01	&	6.22	&	K3  \\
HIP16711	&	03 35 00.5	&	-48 25 11.6	&	< 2.0	&	-5.09	$\pm$	0.01	&	-1.52	$\pm$	0.01	&	8.57	&	K4\\
HIP108870	&	22 03 17.4	&	-56 46 47.3	&	< 2.0	&	-4.80	$\pm$	0.01	&	-1.34	$\pm$	0.07	&	4.69	&	K4\\
HIP85647	&	17 30 11.2	&	-51 38 13.0	&	2.64 $\pm$ 0.36	&	-5.12	$\pm$	0.02	&	-1.53	$\pm$	0.04	&	9.58	&	K7\\
HIP40239	&	08 13 08.5	&	-13 55 01.0	&	4.89 $\pm$ 0.23	&	-5.65	$\pm$	0.25	&	-1.39	$\pm$	0.07	&	9.38	&	M0 \\
HIP29295	&	06 10 34.6	&	-21 51 52.4	&	2.61 $\pm$ 0.41	&	-4.90	$\pm$	0.01	&	-1.51	$\pm$	0.02	&	8.15	&	M1  \\
HIP439	&	00 05 24.2	&	-37 21 25.3	&	< 2.0	&	-5.47	$\pm$	0.01	&	-1.36	$\pm$	0.02	&	8.56	&	M1 \\
HIP42748	&	08 42 44.5	&	+09 33 24.4	&	2.67 $\pm$ 0.72	&	-5.48	$\pm$	0.10	&	-1.39	$\pm$	0.02	&	9.62	&	M1 \\
HIP45908	&	09 21 37.7	&	-60 16 55.1	&	2.32 $\pm$ 0.65	&	-4.94	$\pm$	0.01	&	-1.46	$\pm$	0.02	&	9.49	&	M1 \\
HIP93101	&	18 58 00.1	&	+05 54 29.8	&	2.44 $\pm$ 0.65	&	-4.86	$\pm$	0.01	&	-1.46	$\pm$	0.02	&	9.22	&	M1 \\
HIP93069	&	18 57 30.6	&	-55 59 30.6	&	< 2.0	&	-5.19	$\pm$	0.01	&	--			&	8.86	&	M1\\
HIP23708	&	05 05 47.4	&	-57 33 13.8	&	< 2.0	&	-4.96	$\pm$	0.01	&	-1.42	$\pm$	0.05	&	9.11	&	M2 \\
HIP51317	&	10 28 55.6	&	+00 50 28.0	&	2.28 $\pm$ 0.20	&	-5.16	$\pm$	0.02	&	-1.41	$\pm$	0.01	&	9.65	&	M2 \\
HIP86287	&	17 37 53.3	&	+18 35 29.7	&	2.63 $\pm$ 0.63	&	-5.43	$\pm$	0.01	&	-1.50	$\pm$	0.02	&	9.62	&	M2 \\
HIP88574	&	18 05 07.6	&	-3 01 52.6	&	2.47  $\pm$ 0.29	&	-5.16	$\pm$	0.01	&	-1.51	$\pm$	0.02	&	9.37	&	M2 \\
HIP106440	&	21 33 34.0	&	-49 00 32.0	&	2.09 $\pm$ 0.26	&	-5.23	$\pm$	0.01	&	-1.48	$\pm$	0.02	&	8.66	&	M2 \\
HIP65859	&	13 29 59.7	&	+10 22 38.3	&	2.55 $\pm$ 0.20	&	-4.91	$\pm$	0.01	&	-1.50	$\pm$	0.02	&	9.05	&	M2\\
HIP99701	&	20 13 53.3	&	-45 09 50.4	&	< 2.0	&	-4.84	$\pm$	0.06	&	-1.46	$\pm$	0.02	&	7.97	&	M2\\
HIP67155	&	13 45 42.7	&	+14 53 42.2	&	--	&	-5.04	$\pm$	0.01	&	-1.44	$\pm$	0.01	&	8.46	&	M2\\
HIP85523	&	17 28 39.9	&	-46 53 42.2	&	2.59 $\pm$ 0.34	&	-5.05	$\pm$	0.01	&	-1.22	$\pm$	0.03	&	9.38	&	M3 \\
HIP71253	&	14 34 16.8	&	-12 31 10.7	&	2.83 $\pm$ 0.51	&	-5.51	$\pm$	0.01	&	-1.43	$\pm$	0.04	&	11.32	&	M4  \\
HIP1242	&	00 15 28.1	&	-16 08 01.4	&	4.30 $\pm$ 0.27	&	-5.68	$\pm$	0.21	&	-1.38	$\pm$	0.04	&	11.49	&	M6 \\
HIP53020	&	10 50 52.1	&	+06 48 29.7	&	4.92 $\pm$ 0.22	&	-5.53	$\pm$	0.10	&	-1.43	$\pm$	0.10	&	11.64	&	M6 \\
HIP87937	&	17 57 48.5	&	+04 41 31.0	&	--	&	-5.42	$\pm$	0.01	&	-1.32	$\pm$	0.07	&	9.54	&	M6 \\
HIP92403	&	18 49 49.3	&	-23 50 10.3	&	5.20 $\pm$ 0.91	&	-4.91	$\pm$	0.02	&	-0.98	$\pm$	0.01	&	10.37	&	M6 \\
HIP115332	&	23 21 37.5	&	+17 17 26.1	&	4.94 $\pm$ 0.23	&	-5.13	$\pm$	0.20	&	-1.37	$\pm$	0.05	&	11.70	&	M6 \\
HIP22627	&	04 52 05.7	&	+06 28 35.7	&	3.82 $\pm$ 0.80	&	-5.35	$\pm$	0.04	&	-1.39	$\pm$	0.03	&	11.94	&	M6\\
HIP62452	&	12 47 56.7	&	+09 45 05.2	&	4.30 $\pm$ 0.24	&	-5.63	$\pm$	0.10	&	-1.41	$\pm$	0.01	&	11.39	&	M6\\
HIP113020	&	22 53 16.7	&	-14 15 49.0	&	2.75 $\pm$ 1.03	&	-5.67	$\pm$	0.02	&	-1.47	$\pm$	0.01	&	10.16	&	M6\\
HIP67164	&	13 45 50.7	&	-17 58 05.3	&	2.52 $\pm$ 0.25	&	-5.48	$\pm$	0.04	&	--			&	11.81	&	M6 \\
HIP86214	&	17 37 03.7	&	-44 19 08.7	&	3.42 $\pm$ 0.79	&	-4.99	$\pm$	0.01	&	-1.44	$\pm$	0.05	&	10.94	&	M7\\
\hline                  
\end{tabular}
\end{table*}

\begin{acknowledgements}
This work was supported by FCT/MCTES through national funds and by FEDER - Fundo Europeu de Desenvolvimento Regional through COMPETE2020 - Programa Operacional Competitividade e Internacionaliza\c{c}\~ao by these grants: UID/FIS/04434/2019; PTDC/FIS-AST/32113/2017 \& POCI-01-0145-FEDER-032113; PTDC/FIS-AST/28953/2017 \& POCI-01-0145-FEDER-028953.
  This work was supported by FEDER - Fundo Europeu de Desenvolvimento Regional through COMPETE2020 - Programa Operacional Competitividade e Internacionaliza\c{c}\~ao (POCI-01-0145-FEDER-028987). S.H. acknowledge support by the fellowships PD/BD/128119/2016 funded by FCT (Portugal).
  V.A., S.G.S. and E.D.M. acknowledge support from FCT through Investigador FCT contracts nrs. IF/00650/2015/CP1273/CT0001; IF/00028/2014/CP1215/CT0002; IF/00849/2015/CP1273/CT0003. J. H. C. M. is supported in the form of work contract (DL 57/2016/CP1364/CT0007) funded by national funds through FCT.
 M.O. acknowledges research funding from the Deutsche Forschungsgemeinschft (DFG, German Research Foundation) - OS 508/1-1. A.F.L and G.M. acknowledges the support by INAF/Frontiera through the "Progetti Premiali" funding scheme of the Italian Ministry of Education, University, and Research. J.I.G.H and RR acknowledge financial support from the Spanish Ministry project MINECO AYA2017-86389-P. J.I.G.H also acknowledges financial support from the Spanish MINECO under the 2013 Ram\'on y Cajal program MINECO RYC-2013-14875. J. H. C. M. is supported in the form of a work contract funded by national funds through Funda\c{c}\~ao para a Ci\^encia e Tecnologia (FCT). B.R.-A. acknowledges the support from CONICYT PAI/Concurso Nacional Inserción en la Academia, Convocatoria 2015 79150050. This work results within the collaboration of the COST Action TD 1308. This research has made use of the SIMBAD database, operated at CDS, Strasbourg, France \cite{wenger2000simbad} and has made use of the VizieR catalogue access tool, CDS, Strasbourg, France, the IRAF facility and the VALD3 database. This research has made use of the NASA Exoplanet Archive, which is operated by the California Institute of Technology, under contract with the National Aeronautics and Space Administration under the Exoplanet Exploration Program. This work has made use of data from the European Space Agency (ESA) mission {\it Gaia} (\url{https://www.cosmos.esa.int/gaia}), processed by the {\it Gaia} Data Processing and Analysis Consortium (DPAC, \url{https://www.cosmos.esa.int/web/gaia/dpac/consortium}). Funding for the DPAC has been provided by national institutions, in particular the institutions participating in the {\it Gaia} Multilateral Agreement. We thank the referee for the critical analysis and constructive comments.
      \end{acknowledgements}

%
%

\bibliographystyle{aa}
\bibliography{bt}

 \begin{appendix}

\section{$H_{\alpha}$ Index \label{halpha}}

On top of the $log(R' _{HK})$ index, we also computed, where it was possible for all our stars, an activity index based on the $H_{\alpha}$ line. 
The activity characterization of M dwarfs is made more difficult than for G and K counterparts due to the comparatively lower flux in the wavelength region corresponding to the Ca II H $\&$ K. Fortunately, several alternatives exist in the optical domain. It was demonstrated that an efficient activity indicator is $H_{\alpha}$, which correlates very well with the S index \citep{gomes2011long}. The $H_{\alpha}$ index is a well known and widely used chromospheric indicator \citep{kuerster2003low, 2007A&A...474..293B}. Further to the analysis of the $log(R' _{HK})$ index, we also defined a homogenized $H_{\alpha}$ index similar to \citet{boisse2009stellar} but using a broader central band as proposed by \citet{pasquini1991h} in order to be more sensitive to chromospheric activity contribution \citep{da2014long}. 

\begin{figure}
   \centering
   \includegraphics[width=9cm]{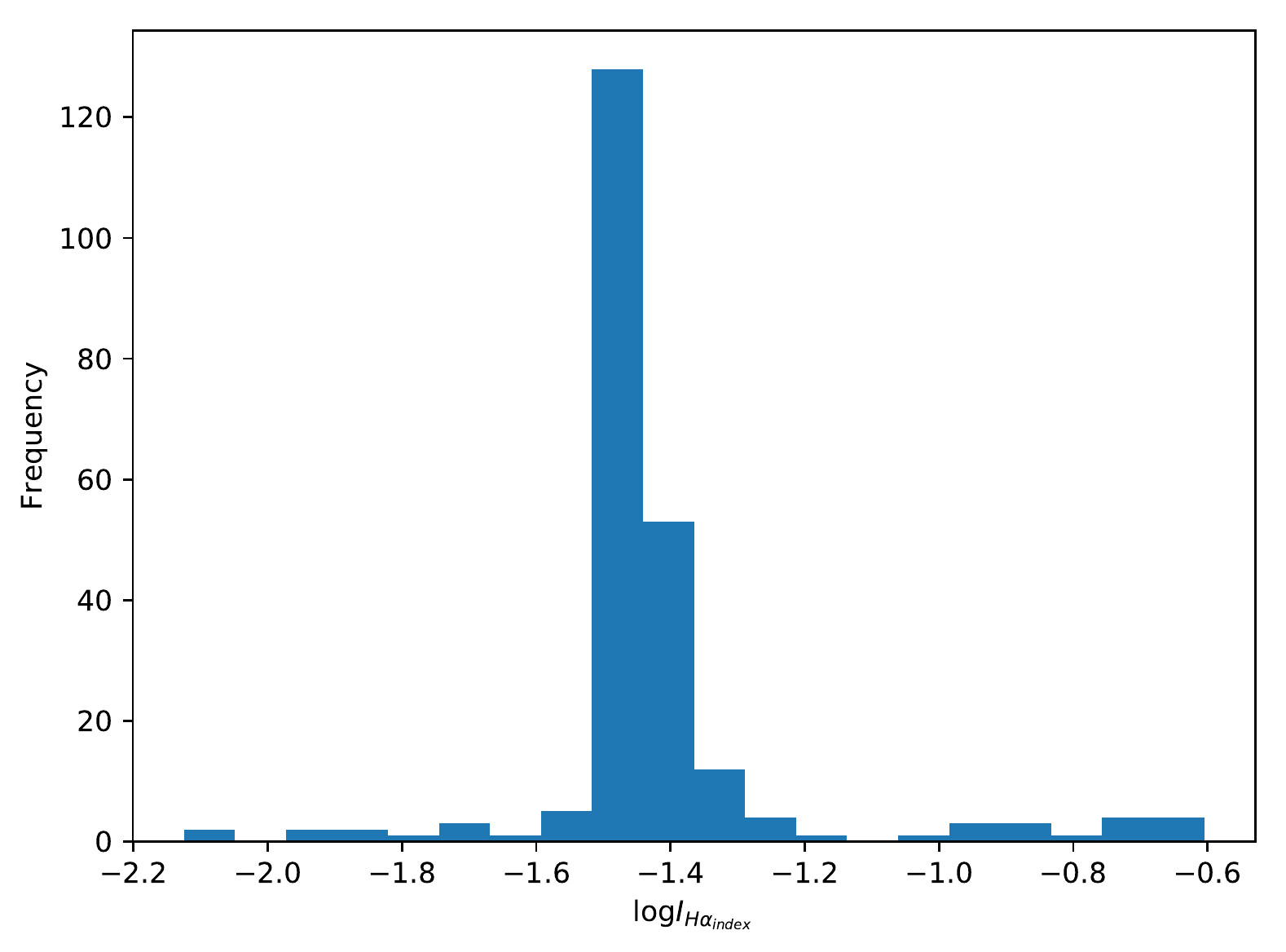}
    
     \caption{Derived $\log I_{H\alpha _{index}}$ distribution for sample stars.}
         \label{ha-histo}
   \end{figure}

We computed the index as described in \citet{boisse2009stellar}:

 \begin{equation}
       H\alpha_{index}  = \frac{F_{H\alpha}}{F_{1} + F_{2}},
  \end{equation}

with $F_{H\alpha}$ sampling the $H_{\alpha}$ line, and the $F_{1}$ and $F_{2}$ as the continuum on both sides of the line. The $H_{\alpha}$ line is centered on 6562.808  $\AA$ with 1.6 $\AA$ wide and continuum $F_{1}$ and $F_{2}$, integrated over [6545.495, 6556.245 $\AA$] and [6575.934, 6584.684 $\AA$], respectively. Each spectrum is corrected for RV shift by its RV measurement by the CCFs. The error was extracted based on the measurement of flux computed by the pipelines and using error propagation.
The $H\alpha _{index}$ as defined in this way is a good indicator to measure activity variation over time for a star. However, it can not be used to compare stars with different spectral types. There is a significant correlation between $H\alpha _{index}$ and $B-V$ that needs to be corrected in order to make $H\alpha _{index}$ compatible between different stars \citep{cincunegui2007halpha}. Here we follow a calibration extracted by \citet{da2014long} and extended it for the $B-V > 1.2$ in order to calibrate $H\alpha _{index}$ and determine $\log I_{H\alpha _{index}}$ (see Table~\ref{result}). The average value of our calibrated $\log H_{\alpha}$ index, $\log I_{H\alpha _{index}}$ is -1.460. As seen from Fig.~\ref{ha-histo}, most stars have values close to -1.460, with only 15 stars having values above -1.150. We note that 14 stars out of these “highly active stars” had already been identified and screened out using the $log(R' _{HK})$, which shows the strong correlation between the two indicators.

\section{Tables.}

\renewcommand{\thefootnote}{\fnsymbol{footnote}}
\longtab[1]{

\begin{longtable}{cccccccc}

\caption{ \label{result} Derived parameters for the first sample stars} \\
\hline\hline

  Target\footnotemark[1]  &  $\textit{v sin i}$ [km $s^{-1}$]  &   $log(R' _{HK})$  &  $\log I_{H\alpha _{index}}$ &   Contamination   &   Program ID \footnotemark[2]  &  $V $ &   Spectral-type \\
 
 \hline
\endfirsthead
\caption{continued.}\\
\hline\hline
 Target & $\textit{v sin i}$ [km $s^{-1}$]  &   $log(R' _{HK})$ &  $\log I_{H\alpha _{index}}$ &   Contamination  &   Program ID &   $V $ &   Spectral-type \\
\hline
\endhead

\hline
\endfoot

HIP93203	&	25.63	$\pm$	0.55	&	-4.74	$\pm$	0.01	&	--			&	N	&	[96],[97],[98]	&	5.27	&	G0 \\
HIP115623	&	80.42	$\pm$	0.93	&	-4.57	$\pm$	0.01	&	-1.37	$\pm$	0.02	&	--	&	[96],[97],[98]	&	4.42	&	G0 \\
HIP23693	&	20.16	$\pm$	0.35	&	-4.33	$\pm$	0.01	&	-1.44	$\pm$	0.03	&	N	&	[88],[89]	&	4.71	&	G0 \\
HIP16852	&	3.36	$\pm$	0.20	&	-5.24	$\pm$	0.01	&	-1.50	$\pm$	0.05	&	N	&	[39],[27]	&	4.30	&	G0 \\
HIP1599	&	< 2.0			&	-4.89	$\pm$	0.01	&	-1.48	$\pm$	0.06	&	N	&	[81],[82]	&	4.23	&	G0 \\
HIP38908	&	< 2.0			&	-4.89	$\pm$	0.01	&	-1.48	$\pm$	0.04	&	N	&	[81],[82]	&	5.59	&	G0 \\
HIP77257	&	2.73	$\pm$	0.21	&	-4.93	$\pm$	0.01	&	-1.48	$\pm$	0.07	&	N	&	[96],[97],[98]	&	4.42	&	G2 \\
HIP22263	&	3.21	$\pm$	0.22	&	-4.52	$\pm$	0.01	&	-2.44	$\pm$	0.01	&	N	&	[2],[5],[40]	&	5.50	&	G2 \\
\end{longtable}
\footnotetext[1]{$\star$ : host confirmed exoplanet}
\footnotetext[2]{ Program ID: 072.C-0488(E): [43], 072.C-0495(A): [65], 073.D-0038(D): [12], 074.B-0639(A): [66], 075.D-0453(A): [72], 076.B-0055(A): [39], 076.C-0279(A): [56], 076.C-0878(B): [48], 076.D-0130(B): [18], 076.D-0130(C): [15], 076.D-0130(E): [53], 077.C-0012(A): [21], 077.C-0364(E): [63], 078.C-0209(B): [34], 078.C-0751(B): [36], 078.C-0833(A): [3], 078.D-0760(B): [60], 079.C-0046(A): [68], 079.C-0521(A): [25], 079.C-0657(C): [17], 079.C-0681(A): [45], 079.D-0075(A): [32], 079.D-0466(A): [29], 080.D-0086(B): [78], 080.D-0086(C): [71], 080.D-0347(A): [26], 081.C-0802(B): [61], 081.C-0802(C): [4], 081.D-0065(E): [16], 082.C-0427(A): [51], 082.C-0427(C): [58], 082.C-0718(B): [22], 082.D-0953(A): [62], 084.C-0229(A): [35], 084.C-0403(A): [24], 085.C-0019(A): [67], 086.C-0145(A): [13], 086.C-0284(A): [33], 087.C-0831(A): [57], 090.C-0421(A): [59], 091.C-0271(A): [64], 092.C-0721(A): [41], 092.D-0207(A): [77], 093.C-0062(A): [55], 093.C-0343(A): [14], 093.C-0409(A): [19], 094.C-0322(A): [50], 094.C-0367(A): [54], 094.D-0596(A): [8], 095.C-0551(A): [1], 096.C-0708(A): [73], 173.C-0606(B): [74], 180.C-0886(A): [75], 183.C-0437(A): [31], 183.C-0972(A): [11], 184.C-0815(C): [28], 188.C-0265(B): [5], 188.C-0265(M): [40], 188.C-0265(O): [2], 191.C-0873: [10], 191.C-0873(A): [6], 192.C-0224: [9], 192.C-0224(B): [42], 192.C-0224(C): [76], 192.C-0224(G): [52], 192.C-0224(H): [44], 192.C-0852(A): [46], 266.D-5655(A): [27], 288.C-5010(A): [37], 289.C-5053(A): [20], 290.C-5196(A): [47], 290.C-5196(B): [23], 380.C-0083(A): [38], 60.A-9036(A): [69], 67.D-0321(B): [70], 68.D-0166(A): [30], 71.C-0498(A): [49], Lagrange: [7], '077.C-0530(A)': [79], '076.D-0103(A)': [80], '198.C-0836(A)': [81], '097.C-0021(A)': [82], '60.A-9700(G)': [83], '073.D-0038(C)': [84], '074.C-0012(A)': [85], '074.C-0012(B)': [86], '196.C-1006(A)': [87], '184.C-0815(A)': [88], '184.C-0815(F)': [89], '072.D-0707(A)': [90], '088.C-0513(B)': [91], '495.L-0963(A)': [92], '096.C-0460(A)': [93], '082.C-0315(A)': [94], '099.C-0880(A)': [95], '97.C-0561(A)': [96], '97.C-0561(B)': [97], $A33TAC\_7$: [98] }

}

\longtab[2]{
\begin{longtable}{ccccccc}
\caption{\label{stellar-param-table} Stellar Parameters}\\

\hline\hline
  Target & R.A [hh:mm:ss] & Dec [$^{\circ}$:$^\prime$:$''$] & \teff\ [K]  &   $\log g$ [cm $s^{-2}$]  &  $[Fe/H]$ &   $\xi_{\mathrm{t}}$ [km $s^{-1}$] \\
\hline
\endfirsthead
\caption{continued.}\\
\hline\hline
  Target & R.A [hh:mm:ss] & Dec [$^{\circ}$:$^\prime$:$''$] & \teff\ [K]  &   $\log g$ [cm $s^{-2}$]  &  $[Fe/H]$ &   $\xi_{\mathrm{t}}$ [km $s^{-1}$]  \\
\hline
\endhead

\hline
\endfoot

HIP439	&	00 05 24.1	&	-37 21 25.3	&	3773 $\pm$ 92	&	--	&	-0.51 $\pm$ 0.08	&	-- \\
HIP1242	&	00 15 28.0	&	-16 08 01.3	&	3186 $\pm$ 92	&	4.681	&	-0.33 $\pm$ 0.05	&	-- \\
HIP1421	&	00 17 47.6	&	+01 41 19.3	&	5054 $\pm$ 29	&	2.98  $\pm$ 0.069	&	0.046  $\pm$ 0.024	&	1.396  $\pm$ 0.029 \\
HIP1599	&	00 20 01.9	&	-64 52 39.4	&	5962 $\pm$ 18	&	4.5 $\pm$ 0.03	&	-0.17 $\pm$ 0.01	&	1.08 $\pm$ 0.03 \\
HIP2540	&	00 32 14.7	&	-63 05 27.6	&	3924	&	4.105	&	--	&	-- \\
HIP3765	&	00 48 22.5	&	+05 17 00.2	&	4962 $\pm$ 49	&	4.433  $\pm$ 0.112	&	-0.338  $\pm$ 0.024	&	0.506  $\pm$ 0.141 \\
HIP3834	&	00 49 13.9	&	-24 08 11.5	&	5003 $\pm$ 43	&	3.279  $\pm$ 0.097	&	0.064  $\pm$ 0.032	&	1.098  $\pm$ 0.05 \\
HIP4569	&	00 58 27.8	&	-27 51 25.1	&	3402 $\pm$ 92	&	4.036	&	0.00 $\pm$ 0.05	&	-- \\
\end{longtable}

}


%
\longtab[3]{
\tiny
\begin{landscape}
\begin{longtable}{ccccccccccc}
\caption{\label{chemichal1} Chemical abundances relative to the Sun}\\
\hline\hline
star & [NaI/H] & [MgI/H] & [AlI/H] & [SiI/H] & [CaI/H] & [ScII/H] & [TiI/H] & [TiII/H] & [CrI/H]& [NiI/H] \\
\hline
\endfirsthead
\caption{continued.}\\
\hline\hline
star & [NaI/H] & [MgI/H] & [AlI/H] & [SiI/H] & [CaI/H] & [ScII/H] & [TiI/H] & [TiII/H] & [CrI/H]& [NiI/H] \\
\hline
\endhead
\hline
\endfoot
HIP1421	&	0.150 $\pm$ 0.017	&	0.107 $\pm$ 0.056	&	0.066 $\pm$ 0.022	&	0.119 $\pm$ 0.063	&	0.031 $\pm$ 0.056	&	0.057 $\pm$ 0.021	&	0.032 $\pm$ 0.070	&	0.016 $\pm$ 0.056	&	0.013 $\pm$ 0.042	&	0.007 $\pm$ 0.044 \\
HIP1599	&	-0.250 $\pm$ 0.021	&	-0.146 $\pm$ 0.045	&	-0.192 $\pm$ 0.031	&	-0.180 $\pm$ 0.025	&	-0.208 $\pm$ 0.063	&	-0.137 $\pm$ 0.024	&	-0.155 $\pm$ 0.030	&	-0.166 $\pm$ 0.015	&	-0.230 $\pm$ 0.034	&	-0.233 $\pm$ 0.014 \\
HIP3765	&	-0.347 $\pm$ 0.013	&	-0.183 $\pm$ 0.081	&	-0.217 $\pm$ 0.017	&	-0.261 $\pm$ 0.071	&	-0.170 $\pm$ 0.147	&	-0.309 $\pm$ 0.059	&	-0.149 $\pm$ 0.074	&	-0.385 $\pm$ 0.029	&	-0.282 $\pm$ 0.073	&	-0.352 $\pm$ 0.038 \\
HIP3834	&	0.015 $\pm$ 0.040	&	0.117 $\pm$ 0.073	&	0.105 $\pm$ 0.026	&	0.104 $\pm$ 0.067	&	0.078 $\pm$ 0.153	&	0.064 $\pm$ 0.026	&	0.100 $\pm$ 0.097	&	0.025 $\pm$ 0.042	&	0.058 $\pm$ 0.070	&	0.044 $\pm$ 0.053 \\
HIP4801	&	0.359 $\pm$ 0.052	&	0.310 $\pm$ 0.098	&	0.404 $\pm$ 0.035	&	0.316 $\pm$ 0.157	&	0.081 $\pm$ 0.066	&	0.178 $\pm$ 0.083	&	0.172 $\pm$ 0.136	&	0.042 $\pm$ 0.113	&	0.090 $\pm$ 0.061	&	0.199 $\pm$ 0.101 \\
HIP6605	&	-0.169 $\pm$ 0.045	&	0.016 $\pm$ 0.044	&	-0.038 $\pm$ 0.021	&	-0.090 $\pm$ 0.062	&	-0.103 $\pm$ 0.053	&	-0.114 $\pm$ 0.030	&	-0.084 $\pm$ 0.079	&	-0.134 $\pm$ 0.039	&	-0.189 $\pm$ 0.029	&	-0.178 $\pm$ 0.037 \\
HIP6714	&	0.011 $\pm$ 0.021	&	-0.008 $\pm$ 0.080	&	0.082 $\pm$ 0.052	&	0.082 $\pm$ 0.067	&	0.062 $\pm$ 0.039	&	0.005 $\pm$ 0.023	&	0.069 $\pm$ 0.121	&	-0.058 $\pm$ 0.069	&	0.045 $\pm$ 0.037	&	0.001 $\pm$ 0.040 \\
HIP8102	&	-0.533 $\pm$ 0.028	&	-0.257 $\pm$ 0.072	&	-0.296 $\pm$ 0.027	&	-0.407 $\pm$ 0.020	&	-0.364 $\pm$ 0.095	&	-0.405 $\pm$ 0.046	&	-0.265 $\pm$ 0.036	&	-0.397 $\pm$ 0.031	&	-0.495 $\pm$ 0.044	&	-0.538 $\pm$ 0.023 \\
\end{longtable}
\end{landscape}
}

\longtab[4]{
\tiny
\begin{longtable}{ccccc}
\caption{\label{chemichal2} [X/Fe] abundance ratios for Carbon (X=C) and Oxygen (X=O) with errors in measurements of lines for sample of stars with \teff\ >5200 K}\\
\hline\hline
star & [O/Fe] & errO & [C/Fe] & errC \\ 
\hline
\endfirsthead
\caption{continued.}\\
\hline\hline
star & [O/Fe] & errO & [C/Fe] & errC \\ 
\hline
\endhead
\hline
\endfoot

HIP1599	&	0.069	&	0.090	&	-0.072	&	0.037 \\
HIP8102	&	0.288	&	0.101	&	0.188	&	0.089 \\
HIP10138	&	0.176	&	0.090	&	0.177	&	0.068 \\
HIP13402	&	-0.014	&	0.092	&	-0.024	&	0.077 \\
HIP15330	&	0.054	&	0.132	&	-0.008	&	0.014 \\
HIP15371	&	0.219	&	0.100	&	-0.014	&	0.026 \\
HIP15510	&	99.990	&	0.000	&	0.107	&	0.129 \\
HIP16852	&	0.025	&	0.058	&	-0.048	&	0.031 \\
HIP22263	&	-0.016	&	0.113	&	-0.185	&	0.017 \\
HIP23835	&	0.314	&	0.048	&	0.086	&	0.014 \\
HIP29271	&	0.050	&	0.080	&	-0.050	&	0.015 \\
HIP37606	&	0.016	&	0.057	&	-0.076	&	0.018 \\
HIP37853	&	99.990	&	0.000	&	0.328	&	0.087 \\
HIP38908	&	0.383	&	0.062	&	0.162	&	0.042 \\
HIP40693	&	-0.041	&	0.107	&	-0.034	&	0.031 \\
HIP41926	&	99.990	&	0.000	&	0.117	&	0.135 \\
HIP56452	&	0.184	&	0.109	&	0.150	&	0.174 \\
HIP57443	&	0.200	&	0.103	&	0.040	&	0.084 \\
HIP64924	&	-0.014	&	0.095	&	-0.053	&	0.021 \\
HIP77257	&	0.013	&	0.038	&	-0.037	&	0.024 \\
HIP81300	&	-0.094	&	0.117	&	0.042	&	0.084 \\
HIP83541	&	-0.113	&	0.074	&	-0.093	&	0.049 \\
HIP86796	&	-0.039	&	0.048	&	-0.063	&	0.028 \\
HIP95447	&	-0.153	&	0.063	&	-0.162	&	0.042 \\
HIP98066	&	-0.100	&	0.065	&	-0.125	&	0.027 \\
HIP98767	&	-0.081	&	0.050	&	-0.083	&	0.027 \\
HIP101345	&	0.020	&	0.052	&	-0.063	&	0.014 \\
HIP101916	&	-0.027	&	0.049	&	-0.122	&	0.037 \\
HIP107649	&	-0.114	&	0.081	&	-0.126	&	0.034 \\
HIP110109	&	0.020	&	0.106	&	-0.086	&	0.042 \\
HIP113357	&	-0.243	&	0.048	&	-0.073	&	0.014 \\
HIP114699	&	0.172	&	0.058	&	-0.071	&	0.037 \\

\end{longtable}
}

\longtab[5]{
\tiny
\begin{longtable}{ccccc}
\caption{ 21 Spectroscopic identified binary by \citet{pourbaix2004s} in the sample \label{sp-bin} }\\
\hline\hline
Star & R.A [hh:mm:ss] & Dec [$^{\circ}$:$^\prime$:$''$]  & $V$ & Spectral-type \\ 
\hline
\endfirsthead
\caption{continued.}\\
\hline\hline
Star & R.A [hh:mm:ss] & Dec [$^{\circ}$:$^\prime$:$''$]  & $V$ & Spectral-type \\ 
\hline
\endhead
\hline
\endfoot

HIP4592	&	00 58 45.75	&	-05 52 58.1	&	6.59	&	K0 \\
HIP11964	&	02 34 22.57	&	-43 47 46.9	&	8.70	&	K7Ve \\
HIP17207	&	03 41 10.52	&	+03 36 41.0	&	9.96	&	M0V \\
HIP23221	&	04 59 50.45	&	-10 15 48.0	&	5.38	&	K0IV \\
HIP25282	&	05 24 28.91	&	-00 53 28.8	&	5.08	&	G9III-IV \\
HIP27913	&	05 54 22.98	&	+20 16 34.2	&	4.41	&	G0V \\
HIP29982	&	06 18 40.35	&	+09 02 49.9	&	6.24	&	G5III \\
HIP34608	&	07 10 06.68	&	+21 14 49.1	&	6.43	&	G8V \\
HIP39198	&	08 01 03.40	&	-06 25 09.1	&	6.46	&	G5 \\
HIP45170	&	09 12 17.55	&	+14 59 45.7	&	6.51	&	G9V \\
HIP55505	&	11 22 05.29	&	-24 46 39.8	&	9.41	&	K4-5 \\
HIP73182	&	14 57 26.54	&	-21 24 41.5	&	8.33	&	M1.5V \\
HIP82817	&	16 55 28.75	&	-08 20 10.8	&	9.02	&	M3Ve \\
HIP86400	&	17 39 16.92	&	+03 33 18.9	&	6.52	&	K3V \\
HIP90729	&	18 30 41.65	&	+20 48 54.2	&	6.50	&	G8IV \\
HIP91751	&	18 42 36.12	&	-07 04 25.2	&	6.15	&	K1III \\
HIP95066	&	19 20 32.91	&	-05 24 56.8	&	5.00	&	G8III-IV \\
HIP98351	&	19 58 56.43	&	-30 32 18.2	&	6.28	&	K1III \\
HIP101955	&	20 39 37.71	&	+04 58 19.3	&	7.89	&	K5V \\
HIP107095	&	21 41 32.86	&	-14 02 51.4	&	5.18	&	G2IV \\
HIP111802	&	22 38 45.58	&	-20 37 16.1	&	9.09	&	dM1e \\

\end{longtable}
}

\end{appendix}
\end{document}